\documentclass[11pt,letterpaper,twoside,reqno]{amsart}

\usepackage{amsmath}
\usepackage{amssymb}
\usepackage{bm}
\usepackage{mathrsfs}
\usepackage[pdftex]{graphicx}

\usepackage{color}

\usepackage{url}

\usepackage{supertabular}

\newtheorem{thm}{Theorem}
\newtheorem{cor}[thm]{Corollary}
\newtheorem{lemma}[thm]{Lemma}
\newtheorem{prop}[thm]{Proposition}

\newtheorem{defn}[thm]{Definition}

\numberwithin{equation}{section}

\DeclareMathAlphabet{\mathsfsl}{OT1}{cmss}{m}{sl}

\newcommand{\term}{\emph}

	{\makebox{\phantom{}} \\ %
		\textsc{Input:}\begin{itemize}}%
	{\end{itemize}}

	{\textsc{Output:}\begin{itemize}}%
	{\end{itemize}}

	{\textsc{Procedure:}\begin{enumerate}}%
	{\end{enumerate}}

\renewcommand{\phi}{\varphi}

\newcommand{\eps}{\varepsilon}

\newcommand{\onevct}{\mathbf{e}}

\newcommand{\ZZ}{\mathbb{Z}}

\newcommand{\abs}[1]{\left\vert {#1} \right\vert}

\newcommand{\Prob}{\operatorname{\mathbb{P}}}
\newcommand{\Expect}{\operatorname{\mathbb{E}}}
\newcommand{\E}{\Expect}

\newcommand{\mtx}[1]{\bm{#1}}

\newcommand{\norm}[1]{\left\Vert {#1} \right\Vert}

\newcommand{\smnorm}[2]{{\bigl\Vert {#2} \bigr\Vert}_{#1}}

\newcommand{\pnorm}[2]{\norm{#2}_{#1}}

\newcommand{\opt}{\mathrm{opt}}

\newcommand{\pest}{p_{\mathrm{est}}}

\newcommand{\Fee}{\mtx{\Phi}}

\newcommand{\poly}{\operatorname{poly}}
\newcommand{\polylog}{\operatorname{polylog}}

\newtheorem{claim}[thm]{Claim}

\newcommand\rtp{{\otimes_r}}

\newtheorem{inv}[thm]{Invariant}
\newtheorem{cond}[thm]{Condition}

\usepackage{rotating}

\usepackage{fullpage}

\begin{document}

\title[Algorithmic Dimension Reduction] {Algorithmic Linear Dimension
  Reduction \\ in the $\ell_1$ Norm for Sparse Vectors}

\author[Gilbert, Strauss, Tropp, Vershynin]{A.\ C.\ Gilbert, M.\ J.\ Strauss, J.\ A.\ Tropp, and R.\ Vershynin}

\thanks{Gilbert and Tropp are with the Department of Mathematics, The
  University of Michigan at Ann Arbor, 2074 East Hall, 530 Church St.,
  Ann Arbor, MI 48109-1043.  E-mail:
  \{\url{annacg}$|$\url{jtropp}\}\url{@umich.edu}.  Strauss is with
  the Department of Mathematics and the Department of Electrical
  Engineering and Computer Science, The University of Michigan at Ann
  Arbor.  E-mail: \url{martinjs@umich.edu}.  Vershynin is with the
  Department of Mathematics, The University of California at Davis,
  Davis, CA 95616. E-mail: \url{vershynin@math.ucdavis.edu}.  ACG is
  an Alfred P. Sloan Research Fellow and has been supported in part by
  NSF DMS 0354600.  MJS has been supported in part by NSF DMS 0354600
  and NSF DMS 0510203.  JAT has been supported by NSF DMS 0503299.
  ACG, MJS, and JAT have been partially supported by DARPA ONR
  N66001-06-1-2011.  RV is an Alfred P.\ Sloan Research Fellow. He was
  also partially supported by the NSF grant DMS 0401032.}

\date{\today}

\begin{abstract}
  We can recover approximately a sparse signal with limited noise,
  {\em i.e}, a vector of length $d$ with at least $d-m$ zeros or
  near-zeros, using little more than $m\log(d)$ nonadaptive linear
  measurements rather than the $d$ measurements needed to recover an
  arbitrary signal of length $d$.  Several research communities are
  interested in techniques for measuring and recovering such signals
  and a variety of approaches have been proposed.  We focus on two
  important properties of such algorithms.
\begin{itemize}
\item {\bf Uniformity.}  A single measurement matrix should
  work simultaneously for all signals.
\item {\bf Computational Efficiency.}  The time to recover such an
  $m$-sparse signal should be close to the obvious lower bound,
  $m\log(d/m)$.
\end{itemize}
  To date, algorithms for signal recovery that provide a uniform measurement
  matrix with approximately the optimal number of measurements, such
  as first proposed by Donoho
  and his collaborators, and, separately, by Cand{\`e}s and Tao, are based on
  linear programming and require time $\poly(d)$ instead of
  $m\polylog(d)$.  On the other hand, fast decoding algorithms to date
  from the Theoretical Computer Science and Database communities fail
  with probability at least $1/\poly(d)$, whereas we need failure probability
  no more than around $1/d^m$ to achieve a uniform failure
  guarantee.

  This paper develops a new method for recovering $m$-sparse signals
  that is simultaneously uniform and quick.  We
  present a reconstruction algorithm whose run time, $O( m \log^2(m)
  \log^2(d))$, is {\em sublinear} in the length $d$ of the signal. The
  reconstruction error is within a logarithmic factor (in $m$) of the
  optimal $m$-term approximation error in $\ell_1$.  In particular,
  the algorithm recovers $m$-sparse signals perfectly and noisy signals
  are recovered with polylogarithmic distortion.  Our algorithm makes 
   $O(m \log^2 (d))$ measurements, which is within a logarithmic
  factor of optimal.  We also present a small-space implementation of
  the algorithm.

  These sketching techniques and the corresponding reconstruction
  algorithms provide an algorithmic dimension reduction in the
  $\ell_1$ norm.  In particular, vectors of support $m$ in dimension
  $d$ can be linearly embedded into $O(m \log^2 d)$ dimensions with
  polylogarithmic distortion.  We can reconstruct a vector from its
  low-dimensional sketch in time $O(m \log^2(m) \log^2(d))$.
  Furthermore, this reconstruction is stable and robust under small
  perturbations.
\end{abstract}


\maketitle
\thispagestyle{empty}

\section{Introduction}

We say that a metric space $(X,d_X)$ embeds into a metric space
$(Y,d_Y)$ with distortion $D$ if there are positive numbers
$A,B$ such that $B/A \le D$ and a map $\Fee: X \to Y$ such that
\begin{equation}	\label{bi-Lipschitz}
   A \; d_X(x,y) \le d_Y(\Fee(x), \Fee(y)) \le B \; d_X(x,y) 
        \ \ \ \text{for all $x,y \in X$}.
\end{equation}
A fundamental problem is to understand when a finite metric space,
which is isometrically embedded in some normed space $X$, admits a
dimension reduction; {\em i.e.}, when we can embed it in an appropriate
normed space $Y$ of low dimension.  Dimension reduction techniques
enjoy a wide variety of algorithmic applications, including data
stream computations~\cite{CM03:Whats-Hot,GGIKMS02:Fast-Small-Space}
and approximate searching for nearest
neighbors~\cite{IN05:Nearest-Neighbor} (to cite just a few).  The
dimension reduction result of Johnson and Lindenstrauss~\cite{JL84} is a
fundamental one.  It states that any set of $N$ points in $\ell_2$ can
be embedded in $\ell_2^n$ with distortion $(1+\epsilon)$ and where the
dimension $n = O(\log(N)/\epsilon^2)$.

A similar problem in the $\ell_1$ space had been a longstanding open
problem; Brinkman and Charikar~\cite{BC03:impossible} solved it in the
negative (see
another example in~\cite{NL04:Diamond}).  There exists a set of $N$
points in $\ell_1$ such that any embedding of it into $\ell_1^n$ with
distortion $D$ requires $n = N^{\Omega(1/D^2)}$ dimensions. Thus, a
dimension reduction in $\ell_1$ norm with constant distortion is not
possible. However, it is well known how to do such a dimension
reduction with a logarithmic distortion.  One first embeds any
$N$-point metric space into $\ell_2$ with distortion $O(\log N)$ using
Bourgain's theorem \cite{Bo}, then does dimension reduction
in $\ell_2$ using Johnson-Lindenstrauss result~\cite{JL84}, and
finally embeds $\ell_2^n$ into $\ell_1^{2n}$ with constant
distortion using Kashin's theorem (\cite{K}, see Corollary~2.4 in
\cite{Pi}).  For linear embeddings $\Fee$, even distortions of
polylogarithmic order are not achievable. Indeed, Charikar and Sahai
\cite{CS02:Dim-Red} give an example for which any linear embedding
into $\ell_1^n$ incurs a distortion $\Omega(\sqrt{N/n})$.

Two fundamental questions arise from the previous discussion.
\begin{enumerate}
 \item What are spaces for which a dimension reduction in the $\ell_1$
   norm is possible with constant distortion?
 \item What are spaces for which a linear dimension reduction
 in the $\ell_1$ norm is possible with constant or polylogarithmic distortion?
\end{enumerate}

One important space which addresses question (2) positively consists of
all vectors of small support.  Charikar and Sahai~\cite{CS02:Dim-Red}
prove that the space of vectors of support $m$ in dimension $d$ can be
linearly embedded into $\ell_1^n$ with distortion $1+\epsilon$ with
respect to the $\ell_1$ norm, where $n = O((m/\epsilon)^2 \log d)$
(Lemma 1 in~\cite{CS02:Dim-Red}).  They do not, however, give a
reconstruction algorithm for such signals and their particular
embedding does not lend itself to an efficient algorithm.

The main result of our paper in an {\em algorithmic} linear dimension
reduction for the space of vectors of small support.  The algorithm
runs in sublinear time and is stable.
\begin{thm}
Let $Y$ be a set of points in $\mathbb{R}^d$ endowed with the $\ell_1$
norm. Assume that each point has non-zero coordinates in at most $m$
dimensions.  Then these points can be linearly embedded into $\ell_1$
with distortion $O(\log^2(d) \log^3(m))$, using only $O(m \log^2 d)$
dimensions.  Moreover, we can reconstruct a point from its
low-dimensional sketch in time $O(m \log^2(m) \log^2(d))$.
\end{thm}
This dimension reduction reduces the quadratic order of $m$
in~\cite{CS02:Dim-Red} to a linear order.  Our embedding does,
however, incur a distortion of polylogarithmic order.  In return for
this polylogarithmic distortion, we gain an {\em algorithmic linear
  dimension reduction}---there exists a sublinear time algorithm that
can reconstruct every vector of small support from its low-dimensional
sketch.

The space of vectors of support $m$ in dimension $d$ is a
natural and important space as it models closely the space of
compressible signals.  A \term{compressible signal} is a long signal
that can be represented with an amount of information that is small
relative to the length of the signal.  Many classes of $d$-dimensional
signals are compressible, {\em e.g.},
\begin{itemize}
\item The $m$-sparse class $B_0(m)$ consists of signals with at most
  $m$ nonzero entries.
\item For $0 < p < 1$, the weak $\ell_p$ class $B_{\text{weak-} p}(r)$
  contains each signal $f$ whose entries, sorted by decaying
  magnitude, satisfy $\abs{f}_{(i)} \leq r \, i^{-1/p}$.
\end{itemize}
These types of signals are pervasive in applications.  Natural images
are highly compressible, as are audio and speech signals.  Image,
music, and speech compression algorithms and coders are vital pieces
of software in many technologies, from desktop computers to MP3
players.  Many types of automatically-generated signals are also
highly redundant.  For example, the distribution of bytes per source
IP address in a network trace is compressible---just a few source IP
addresses send the majority of the traffic.

One important algorithmic application of our dimension reduction is
the reconstruction of compressible signals.  This paper describes a
method for constructing a random linear operator $\Fee$ that maps each
signal $f$ of length $d$ to a sketch of size $O( m \log^2 d )$.  We
exhibit an algorithm called Chaining Pursuit that, given this sketch
and the matrix $\Fee$, constructs an $m$-term approximation of the
signal with an error that is within a logarithmic factor (in $m$) of
the optimal $m$-term approximation error.  A compressible signal is
well-approximated by an $m$-sparse signal so the output of Chaining
Pursuit is a good approximation to the original signal, in addition to
being a compressed represention of the original signal.  Moreover,
this measurement operator succeeds simultaneously for all signals with
high probability.  In manyof the above application settings, we have
resource-poor encoders which can compute a few random dot products
with the signal but cannot store the entire signal nor take many
measurements of the signal.  The major innovation of this result is to
combine sublinear reconstruction time with stable and robust linear
dimension reduction of all compressible signals.

Let $f_m$ denote the best $m$-term representation for $f$; {\em i.e.},
$f_m$ consists of $f$ restricted to the $m$ positions that have
largest-magnitude coefficients.
\begin{thm} \label{thm:main-result}
With probability at least $(1 - O(d^{-3}))$, the random measurement
operator $\Fee$ has the following property.  Suppose that $f$ is a
$d$-dimensional signal whose best $m$-term approximation with respect
to $\ell_1$ norm is $f_m$.  Given the sketch $V = \Fee f$ of size $O(m
\log^2 (d))$ and the measurement matrix $\Fee$, the Chaining Pursuit
algorithm produces a signal $\widehat{f}$ with at most $m$ nonzero
entries.  The output $\widehat{f}$ satisfies
\begin{equation}   \label{eqn:main-error}
   \| f - \widehat{f}\|_1
	\leq C(1 + \log m) \smnorm{1}{ f - f_m }.
\end{equation}
In particular, if $f_m = f$, then also $\widehat{f} = f$.  The time
cost of the algorithm is $O( m \log^2 (m) \log^2 (d) )$.
\end{thm}
\begin{cor}
  The factor $\log m$ is intrinsic to this
  approach.  However, the proof gives a stronger statement---the
  approximation in the weak-1 norm {\em without} that factor: $\|f -
  \widehat f\|_{\rm weak-1} \leq C \|f - f_m\|_1$.  This follows
  directly from the definition of the weak norm and our proof, below.
\end{cor}
\begin{cor}\label{rem:stable}
  Our argument
  shows that the reconstruction $\widehat f$ is not only stable with
  respect to noise in the signal, as Equation~\eqref{eqn:main-error}
  shows, but also with respect to inaccuracy in the measurements.
  Indeed, a stronger inequality holds.  For every $V$ (not
  necessarily the sketch $\Fee f$ of $f$) if $\widehat f$ is the
  reconstruction from $V$ (not necessarily from $\Fee f$),  we have
\[   \|f_m - \widehat f\|_1 \leq C(1+ \log m)
          \Big(\|f - f_m\|_1 + \|\Fee f - V\|_1 \Big).
\]
\end{cor}

\subsection{Related Work}

The problem of sketching and reconstructing $m$-sparse and
compressible signals has several precedents in the Theoretical
Computer Science literature, especially the paper
\cite{CM03:Whats-Hot} on detecting heavy hitters in nonnegative data
streams and the works
\cite{GGIMS02:Near-Optimal-Sparse,GMS05:Improved-Time} on Fourier
sampling.  More recent papers from Theoretical Computer Science
include \cite{CM05:Towards-Algorithmic,CRTV05:Error-Correction}.
Sparked by the papers \cite{Don04:Compressed-Sensing} and
\cite{CT04:Near-Optimal}, the computational harmonic analysis and
geometric functional analysis communities have produced an enormous
amount of work, including
\cite{CRT04:Exact-Signal,Don05:Neighborly-Polytopes,DT05:Sparse-Nonnegative,CT05:Decoding-Linear,RV05:Geometric-Approach,TG05:Signal-Recovery,MPTJ05:Reconstruction-Subgaussian}.

Most of the previous work has focused on a reconstruction algorithm
that involves linear programming (as first investigated and promoted
by Donoho and his collaborators) or second-order cone programming
\cite{Don04:Compressed-Sensing,CT04:Near-Optimal,CRTV05:Error-Correction}.
The authors of these papers do not report computation times, but they
are expected to be
cubic in the length $d$ of the signal.  This cost is high, since we
are seeking an approximation that involves $O(m)$ terms.  The paper
\cite{TG05:Signal-Recovery} describes another algorithm with running
time of order $O(m^2 d \log d)$, which can be reduced to $O(m d \log
d)$ in certain circumstances.  None of these approaches is comparable
with the sublinear algorithms described here.

There are a few sublinear algorithms available in the literature.  The
Fourier sampling paper \cite{GMS05:Improved-Time} can be viewed as a
small space, sublinear algorithm for signal reconstruction.  Its
primary shortcoming is that the measurements are not uniformly good
for the entire signal class.  The recent work
\cite{CM05:Towards-Algorithmic} proposes some other sublinear
algorithms for reconstructing compressible signals.  Few of these
algorithms offer a uniform guarantee.  The ones that do require more
measurements---$O(m^2 \log d)$ or worse---which means that they are
not sketching the signal as efficiently as possible.  

Table \ref{tab:comparison} compares the major algorithmic
contributions.  Some additional comments on this table may help
clarify the situation.  If the signal is $f$ and the output is
$\widehat{f}$, let $E=E(f)=f-\widehat{f}$ denote the error vector of
the output and let $E_\opt=E_\opt(f)=f-f_m$ denote the error vector
for the optimal output.  Also, let $C_\opt=C_\opt(f)$ denote
$\max_g E_\opt(g)$, where $g$ is the worst possible signal in the
class where $f$ lives.

\begin{sidewaystable}
\centering

\small
\renewcommand{\arraystretch}{2}

\begin{tabular}{|l|l|c|c|c|c|c|c|}
\hline
Approach, References & Signal Class & Uniform & Error bd. & \# Measurements & Storage & Decode time \\
\hline\hline
OMP + Gauss  \cite{TG05:Signal-Recovery} &
	$m$-sparse & No & No error &
	$m \log d$ & $md\log d$ & $m^2 d \log d$ \\

Group testing \cite{CM06:Combinatorial-Algorithms} &
	$m$-sparse & No & No error &
	$m \log^2 d$ & $\log d$ & $m \log^2 d$ \\

$\ell_1$ min. + Gauss &
	$m$-sparse & Yes & No error &
	No closed form & $\Omega(md)$ & LP$(md)$ \\
\cite{Don06:Neighborly-Polytopes,DT05:Sparse-Nonnegative,DT06:Thresholds-Recovery} &
&&&&&\\
	
Group testing \cite{CM06:Combinatorial-Algorithms} &
	$m$-sparse & Yes & No error &
	$m^2 \log^2 d$ & $m \log(d/m)$ & $m^2 \log^2 d$ \\

\hline
Group testing \cite{CM06:Combinatorial-Algorithms} &
	weak $\ell_p$ & Yes & $\smnorm{2}{E}\le C\smnorm{p}{C_\opt}$ &
	 $m^{\frac{3-p}{1-p}} \log^2 d$ & $m^{\frac{2-p}{1-p}}\log d$ & $m^{\frac{4-2p}{1-p}} \log^3 d$ \\

\hline
$\ell_1$ min. + Gauss &
	Arbitrary & Yes & $\smnorm{2}{E}\le m^{-1/2}\smnorm{1}{E_\opt}$ &
	$m \log(d/m)$ & $d \log(d/m)$ & LP$(md)$  \\
\cite{CT04:Near-Optimal,CDD06:Remarks-Compressed} &
&&&&&\\

$\ell_1$ min. + Fourier &
	Arbitrary & Yes & $\smnorm{2}{E}\le m^{-1/2}\smnorm{1}{E_\opt}$ &
	$m \log^4 d$ & $m \log^5 d$ & $d \log d$ (empirical) \\
\cite{CT04:Near-Optimal,RV06:Sparse-Reconstruction,CDD06:Remarks-Compressed} &
&&&&&\\

\hline
\hline
Chaining Pursuit &
	Arbitrary & Yes & $\smnorm{{\rm weak-1}}{E}\le \smnorm{1}{E_\opt}$ &
	$m \log^2 d$ & $d \log^2 d$ & $m \log^2 d$ \\
\cite{GSTV06:ADR} &
	 & & $\smnorm{1}{E}\le \log(m)\smnorm{1}{E_\opt}$ &
    & & \\

\hline
\hline
Fourier sampling &
	Arbitrary & No & $\smnorm{2}{E} \le \smnorm{2}{E_\opt}$ &
	$m \polylog d$ & $m \polylog d$ & $m \polylog d$ \\
\cite{GGIMS02:Near-Optimal-Sparse,GMS05:Improved-Time} &
&&&&&\\

Group testing \cite{CM06:Combinatorial-Algorithms} &
	Arbitrary & No & $\smnorm{2}{E} \le \smnorm{2}{E_\opt}$ &
	$m \log^{5/2} d$ & $\log^2 d$ & $m \log^{5/2} d$ \\
\hline
\end{tabular} \\
\vspace{1pc}
\textbf{Notes:}  Above, LP$(md)$ denotes resources
needed to solve a linear program with $\Theta(md)$ variables, plus
minor overhead.  We suppress big-O notation for legibility.
\vspace{1pc}

\caption{Comparison of algorithmic results for compressed sensing} \label{tab:comparison}

\end{sidewaystable}

\subsection{Organization}
In Section~\ref{sec:sketch}, we provide an overview of determining a
sketch of the signal $f$.  In Section~\ref{sec:smallspace}, we give an
explicit construction of a distribution from which the random linear
map $\Fee$ is drawn.  In Section~\ref{sec:decode}, we detail the
reconstruction algorithm, Chaining Pursuit, and in
Section~\ref{sec:chaining-analysis} we give an analysis of the
algorithm, proving our main result.  In
Section~\ref{sec:embed-analysis} we use our algorithmic analysis to
derive a dimension reduction in the $\ell_1$ norm for sparse vectors.

\section{Sketching the Signal}\label{sec:sketch}

This section describes a linear process for determining a sketch
$V$ of a signal $f$.  Linearity is essential for supporting
additive updates to the signal.  Not only is this property important
for applications, but it arises during the iterative algorithm for
reconstructing the signal from the sketch.  Linearity also makes the
computation of the sketch straightforward, which may be important for
modern applications that involve novel measurement technologies.

\subsection{Overview of sketching process}

We will construct our measurement matrix by combining simple matrices
and ensembles of matrices.  Specifically, we will be interested in
restricting a signal $f$ to a subset $A$ of its $d$ positions and then
restricting to a smaller subset $B\subseteq A$, and it will be convenient
to analyze separately the two stages of restriction.  If $P$ and $Q$ are
0-1 matrices, then each row $P_i$ of $P$ and each row $Q_j$ of $Q$
restricts $f$ to a subset by multiplicative action, $P_if$ and $Q_jf$, and
sequential restrictions are given by $P_iQ_jf=Q_jP_if$.
We use the following notation, similar to~\cite{CM05:Towards-Algorithmic}.
\begin{defn}
Let $P$ be a $p$-by-$d$ matrix and $Q$ a $q$-by-$d$ matrix, with rows
$\{P_i: 0\le i < p\}$ and $\{Q_j: 0\le j < q\}$, respectively.  The
{\em row tensor product} $S=P\rtp Q$ of $P$ and $Q$ is a $pq$-by-$d$ matrix
whose rows are
$\{P_iQ_j: 0\le i < p,\ 0 \le j < q\}$, where $P_iQ_j$ denotes the
componentwise product of two vectors of length $d$.
\end{defn}
The order of the rows in $P\rtp Q$ will not be important in this
paper.  We will sometimes index the rows by the pair $(i,j)$, where
$i$ indexes $P$ and $j$ indexes $Q$, so that $P\rtp Q$ applied to a
vector $x$ yields a $q \times p$ matrix.

Formally, the measurement operator $\Fee$ is a row tensor product
$\Fee=\mtx{B}\rtp\mtx{A}$.  Here, $\mtx{A}$ is a $O(m\log d)\times d$
matrix called the \term{isolation matrix} and $\mtx{B}$ is a $O(\log
d)\times d$ matrix called the \term{bit test matrix}.  The measurement
operator applied to a signal $f$ produces a sketch $V = \Fee f$,
which we can regard as a matrix with dimensions $O(m \log d) \times
O(\log d)$.  Each row of $V$ as a matrix contains the result of
the bit tests applied to a restriction of the signal $f$ by a row of
$A$.  We will refer to each row of the data matrix as a
\term{measurement} of the signal.

\subsection{The isolation matrix}\label{sec:isolation}

The isolation matrix $\mtx{A}$ is a 0-1 matrix with dimensions $O(m
\log d) \times d$ and a hierarchical structure.  Let $a$ be a
sufficiently large constant, to be discussed in the next two sections.
The Chaining Pursuit algorithm makes $K = 1 + \log_a m$ passes (or
``rounds'')  over
the signal, and it requires a different set of measurements for each
pass.  The measurements for the $k$th pass are contained in the
$O( mk\log(d) / 2^k ) \times d$
submatrix $\mtx{A}^{(k)}$.  During the $k$th pass,
the algorithm performs $T_k = O(k \log d)$ trials.  Each trial $t$ is
associated with a further submatrix $\mtx{A}^{(k)}_t$, which has
dimensions $O(m / 2^k) \times d$.  

In summary,
$$
\renewcommand{\arraystretch}{1.5}
\mtx{A} = \left[ \begin{array}{c}
	\mtx{A}^{(1)} \\ \hline \mtx{A}^{(2)} \\ \hline \vdots \\
	\hline \mtx{A}^{(K)} \end{array} \right]
\qquad\text{where}\qquad
\mtx{A}^{(k)} = \left[ \begin{array}{c}
	\mtx{A}^{(k)}_1 \\ \hline \mtx{A}^{(k)}_2 \\ \hline \vdots \\
	\hline \mtx{A}^{(k)}_{T_k} \end{array} \right].
$$

Each trial submatrix $\mtx{A}^{(k)}_{t}$ encodes a random partition of
the $d$ signal positions into $O(m / 2^k)$ subsets.  That is, each
signal position is assigned uniformly at random to one of $O( m / 2^k
)$ subsets.  So the matrix contains a $1$ in the $(i, j)$ position if
the $j$th component of the signal is assigned to subset $i$.
Therefore, the submatrix $\mtx{A}^{(k)}_{t}$ is a 0-1 matrix in which
each column has exactly one 1, {\em e.g.},
$$
\mtx{A}^{(k)}_{t} =
\begin{bmatrix}
0 & 1 & 0 & 0 & 1 & 1 & 0 \\
1 & 0 & 0 & 1 & 0 & 0 & 0 \\
0 & 0 & 1 & 0 & 0 & 0 & 1
\end{bmatrix}.
$$
The trial submatrix can also be viewed as a random linear hash
function from a space of $d$ keys onto a set of $O(m / 2^k)$
buckets.

\subsection{The bit test matrix}

Formally, the matrix $\mtx{B}$ consists of a row $\onevct$ of 1's and
other rows given by a 0-1 matrix $\mtx{B_0}$, which we now describe.
The matrix $\mtx{B}_0$ has dimensions $\log_2 \lceil d \rceil\times
d$.  The $i$th column of $B_0$ is the binary expansion of $i$.
Therefore, the componentwise product
of the $i$th row of $B_0$ with $f$ yields a copy of the signal $f$ with
the components that have bit $i$ equal to one selected and the others
zeroed out.

An example of a bit test matrix with $d = 8$ is
$$
\mtx{B} = \left[ \begin{array}{cccccccc}
1 & 1 & 1 & 1 & 1 & 1 & 1 & 1\\
\hline
0 & 0 & 0 & 0 & 1 & 1 & 1 & 1\\
0 & 0 & 1 & 1 & 0 & 0 & 1 & 1\\
0 & 1 & 0 & 1 & 0 & 1 & 0 & 1
\end{array} \right].
$$

\subsection{Storage costs}

The bit test matrix requires no storage.  The total storage for the
isolation matrix is $O( d \log d )$.  The space required for the
isolation matrix is large, but this space can conceivably be shared
among several instances of the problem.  In
Section~\ref{sec:smallspace}, we give an alternate construction in
which a pseudorandom isolation matrix is regenerated as needed from a
seed of size $m\log^2(d)$; in that construction only the seed needs
to be stored, so the total storage cost is $m\log^2(d)$.

\subsection{Encoding time}

The time cost for measuring a signal is $O( \log^2 (m) \log^2 (d) )$
\emph{per nonzero component}.  This claim follows by observing that a
single column of $\mtx{A}$ contains $O( \log^2 (m) \log (d) )$ nonzero
entries, and we must apply $\mtx{A}$ to each of $O( \log d )$
restrictions of the signal---one for each row of $B$.  Note that this
argument assumes random access to the columns of the isolation matrix.
We will use this encoding time calculation when we determine the time
costs of the Chaining Pursuit algorithm.  In
Section~\ref{sec:smallspace}, we give an alternative construction for
$A$ that reduces the storage requirements at the cost of slightly
increased time requirements.  Nevertheless, in that construction, any
$m$ columns of $A$ can be computed in time $m^{o(1)}$ each, where
$o(1)$ denotes a quantity that tends to 0 as both $m$ and $d$ get
large.  This gives measurement time $m^{o(1)}$ per nonzero component.

\section{Small Space Construction}\label{sec:smallspace}

We now discuss a small space construction of the isolation matrix,
$A$.  The goal is to specify a pseudorandom matrix $A$ from a small
random seed, to avoid the $\Omega(d\log d)$ cost of storing $A$
explicitly.  We then construct entries of $A$ as needed, from the
seed.  If we were to use a standard pseudorandom number generator
without further thought, however, the time to construct an entry of
$A$ might be $\Omega(m)$, compared with $O(1)$ for a matrix that is
fully random and explicitly stored.  We will give a construction that
addresses both of these concerns.

As discussed in Section~\ref{sec:isolation}, the matrix $A$ consists
of $\polylog(d)$ submatrices that are random partitions of the $d$
signal positions into $O(m_k)$ subsets.  In this section, we give a
new construction for each submatrix; the submatrices fit together to
form $A$ in the same way as in Section~\ref{sec:isolation}.  We will
see from the analysis in Section~\ref{sec:chaining-analysis}, the
partition map of each random submatrix need only be $m_k$-wise
independent; full independence is not needed as we need only control
the allocation of $m_k$ spikes into measurements in each submatrix.
It follows that we need only construct a family of $d$ random
variables that are $m_k$-wise independent and take values in
$\{0,\ldots,r-1\}$ for any given $r\le d$.  Our goal is to reduce the
storage cost from $O(d\log d)$ to $m\polylog(d)$ without unduly
increasing the computation time.  It will require time $\Omega(m)$ to
compute the value of any {\em single} entry in the matrix, but we will
be able to compute any submatrix of $m$ columns (which is all zeros
except for one 1 per column) in {\em total} time $m\polylog(d)$.  That
is, the values of any $m$ random variables can be computed in time
$m\polylog(d)$.  As in Theorem~\ref{thm:isolation-matrix} below, our
construction will be allowed to fail with probability $1/d^3$, which
will be the case.  (Note that success probability $1-e^{-c m\log d}$
is not required.)  Our construction combines several known
constructions from~\cite{AHU,CLRS}.  For completeness, we sketch
details.

\subsection{Requirements}

To ease notation, we consider only the case of $m_k=m$.
Our goal is to construct a function
$f_s:\{0,\ldots,d-1\}\to\{0,\ldots,r-1\}$, where $s$ is a random
seed.  The construction should ``succeed'' with probability at least
$1-1/d^3$; the remaining requirements only need to hold if the
construction succeeds.  The function should be uniform and $m$-wise
independent, meaning, for any $m$ distinct positions $0\le
i_1,\ldots,i_m<d$ and any $m$ targets $t_1,\ldots,t_m$, we have
\[  \Prob_s( \forall j\>f_s(i_j)=t_j)=r^{-m}, 
\]
though the distribution on $m+1$ random variables may otherwise be
arbitrary.  Finally, given any list $A$ of $m$ positions, we need to
be able to compute $\{f(j):j\in A\}$ in time $m\polylog(d)$.

\subsection{Construction}

Let $s = (s_0, s_1, \dots, s_K)$ be a sequence of $K \leq O(\log d)$
independent, identically distributed random bits.  Let $p$ be a prime
with $p \geq 2r$ and $d \leq p \leq \poly(d)$.  Define the map $g_s^k:
\ZZ_p\to\ZZ_p$ which uses the $k$th element $s_k$ from
the seed $s$ and maps $j \in \ZZ_p$ uniformly at random to a point
$g_s^k(j) \in \ZZ_p$.  The map $g_s^k$ is a random polynomial of
degree $m-1$ over the field with $p$ elements.  If 
\[   0 \leq g_s^o(j) < r\lfloor p/r\rfloor,
\]
where $r\lfloor p/r\rfloor$ represents the largest multiple of $r$
that is at most $p$, then define
\[  f_s(j) = \lfloor g^0_s(j)  r / p\rfloor = h(g^0_s(j)).
\]
The function $h:\{0,\ldots,r\lfloor p/r\rfloor-1\}\to\{0,\ldots,r-1\}$
is a function that is exactly $\lfloor p/r\rfloor$-to-1.  If
$g^0_s(j)>r\lfloor p/r\rfloor$, then we map $j$ to $\mathbb{Z}_p$ by
$g^1_s(j)$, that is independent of $g^0_s$ and identically
distributed.  We repeat the process until $j$ gets mapped to
$\{0,\ldots,r\lfloor p/r\rfloor-1\}$ or we exhaust $K\le O(\log d)$
repetitions.  For computational reasons, for each $k$, we will compute
$g^k_s$ at once on all values in a list $A$ of $m$ values and we will
write $g^k_s(A)$ for the list $\{ g^k_s(j) | j\in A\}$.
Figure~\ref{fig:hashtop} gives a formal algorithm.
\begin{figure}[h!]
\caption{Top-level description of $m$-wise independent random
  variables.}\label{fig:hashtop}
\begin{center}
\tt
\begin{tabular}{|p{.9\textwidth}|}
\hline
\centerline{\textbf{Algorithm: Hashing}} \\
Parameters: $m,r,d,K$\\
Input: List $A$ of $m$ values in $\mathbb{Z}_d$; pseudorandom seed $s$.\\
Output: List $B$ of $m$ values in $\{0,\ldots,r-1\}$, representing
     $\langle f_s(j):j\in A\rangle$.\\
Compute $g^k_s(A)$ for $k=0,1,2,\ldots,K-1$.\\
If for some $j\in A$, for all $k<K$, we have $g^k_s(j)\ge r\lfloor
p/r\rfloor$, then FAIL\\
For $j\in A$\\
\hspace{2pc} $k_j = \min\{k: g^k_s(j) < r\lfloor p/r\rfloor\}$\\
\hspace{2pc} $f_s(j) = g^{k_j}_s(j)$.\\
\hline
\end{tabular}
\vspace{1pc}
\end{center}
\end{figure}

\subsection{Correctness}

\begin{lemma}
Our construction of $f_s:\{0,\ldots,d-1\} \to \{0, \ldots, r-1\}$
produces a uniform $m$-wise independent partition with probability at
least $1 - d^{-3}$.
\end{lemma}
\begin{proof}
The proof of the correctness of our construction is a standard
argument, which we sketch for completeness.  First, there is a prime
$p$ with $d \leq p \leq \poly(d)$.  Next, let us consider the
construction of $g_s^k$.  Because the definition of $g_s^k$ is
independent of $k$, we drop the $s$ and $k$ and write $g$ for
simplicity.  Because $g$ is a random polynomial of degree $m-1$ over
the field of $p$ elements, we can view the construction of $g$ as the
multiplication of a vector $c$ of length $m$ (the coefficients of $g$)
by the Vandermonde matrix
\[
V=\begin{pmatrix}
1 & 1 &  1  &  1  & \cdots & 1\\
0 & 1 &  2  &  3  & \cdots & p - 1\\
0 & 1 & 2^2 & 3^2 & \cdots & (p - 1)^2\\
\vdots
\end{pmatrix}.
\]
Thus we obtain $cV=(g(0), g(1), \ldots, g(p-1))$.  If $A$ is a list of
$m$ positions, then $g(A)$ is $cV_A$, where $V_A$ is the submatrix of
$V$ gotten by selecting columns according to $A$.  Since $V$ is a
square vandermonde matrix over a field, it is invertible.  It follows
that, as $c$ varies, $cV_A$ varies over all of $\mathbb{Z}_p^m$,
hitting each element exactly once.

Next, $g(j)\ge r\lfloor p/r\rfloor$ with probability at most $r/p\le
1/2$.  It follows that, for some $k< K$, we have, with probability at
least $1-2^{-K}$, that $g(j)<r\lfloor p/r\rfloor$.  For sufficiently
large $K\le O(\log d)$, the probability is at least $1-1/d^4$.  Taking
a union bound over all $d$ possible $j$'s, the construction succeeds
with probability at least $1-1/d^3$.

It is easy to check that, by construction, $f_s(A)$ is uniform on
$\{0,\ldots,r-1\}^m$ conditioned on the construction succeeding.
\end{proof}

\subsection{Efficiency}

\begin{lemma}
Given an arbitrary set $A$ of $m$ positions in $\ZZ_p$ and a degree
$m-1$ polynomial $g$, we can evaluate $g$ on $A$ in time $O(m
\poly\log(d))$.
\end{lemma}
\begin{proof}
Evaluating $g$ on the set $A$ is known as the multipoint polynomial
evaluation (MPE) problem.  We recall that the MPE problem can be
reduced to $\polylog(m)$ polynomial multiplications~\cite{AHU} and
that we can multiply polynomials efficiently using the FFT algorithm.
We observe that the time to multiply polynomials in $m \polylog(d)$ as
we may multiply polynomials by convolving their coefficients (via the
FFT algorithm) over $\mathbb{C}$ and then quantizing and reducing
modulo $p$ the result.  We note that arithmetic modulo $p$ take time
at most $\polylog(d)$.

Let us now review the MPE problem.  Recall that we wish to evaluate
$g$ on the set $A$, $g(A)$.  The evaluation of $g(x)$ at some point $x
= t$ is equivalent to finding $g$ mod $(x-t)$, since we can write
$g(x)=q(x)(x-t)+r$ by the division theorem.  To compute the quotients
$g \mod (x-a$ for each $a \in A$, let us assume that $|A|$ is a power
of 2 (padding if necessary), then we order $A=\{a_i\}$ arbitrarily and
form a binary tree in which the $k$'th node at depth $j$ corresponds
to the subset $A_{j,k}=\{a_i:km/2^j\le i<(k+1)m/2^j\}\subseteq A$.
Once we have formed the binary tree, we compute the polynomials
$p_{j,k}(x)=\prod_{i\in A_{j,k}}(x-a_i)$ at each node.  We also define
$g_{j,k}$ at each node by $g_{j,k}=g\bmod p_{j,k}$.  Our goal is to
compute $g_{\lg m,k}=g \bmod p_{\lg m,k}$ for all $k$, {\em i.e.},
reduce $g$ modulo each polynomial in a leaf of the tree.  To do this,
we start with $g = g_{0,0} = g\bmod p_{0,0}$, {\em i.e.}, $g$ mod the
root polynomial.  {From} $g_{j,k}$ we form the two children,
$g_{j+1,2k}=g_{j,k}\bmod p_{j+1,2k}$ and $g_{j+1,2k+1}=g_{j,k}\bmod
p_{j+1,2k+1}$.  Note that, at depth $j$, we have $2^j$ polynomials
$g_{j,k}$ of degree $m/2^j-1$ and $p_{j,k}$ of degree $m/2^j$.

We form the tree of $p_{j,k}$'s in a straightforward fashion, using
the FFT algorithm to multiply polynomials.  Multiplying a pair of
polynomials at depth $j$ takes time $m/2^j\polylog(d)$ and there are
$O(2^j)$ such problems, for total time $m\polylog(d)$ at depth $j$,
and total time $m\polylog(d)$ in aggregrate over all $O(\log m)$
levels.

It remains to show how to reduce a polynomial $g$ of degree $2n-1$ by
a polynomial $q$ of degree $n$ in time $n\polylog(d)$.  First, we
reduce $x^{n-1+2^k}$ for all $k=0,1,2,\ldots,\lg(n)$.  Suppose we have
done the reduction for for
$x^{n},x^{n+1},x^{n+3},x^{n+7}\ldots,x^{n-1+2^{k-1}}$.  We claim that
we can then reduce any polynomial $h$ of degree $n-1+2^k$ by $q$ in
time $n\polylog(d)$.  To see this, write 
\[  h(x)=x^{n-1+2^{k-1}}h'(x) + h''(x), 
\]
where $h'$ has degree $2^{k-1}$ and $h''$ has degree $n-1+2^{k-1}$.
Then, multiply $x^{n+2^{k-1}-1}$ mod $q$ by $h'$ and obtain a
polynomial $h$ of degree $n-1+2^{k-1}$ which we add to $h''$.  This
reduces the problem for a polynomial of degree $n-1+2^k$ to a
polynomial of degree $n-1+2^{k-1}$, in time $n\polylog(d)$.  Let us
perform this reduction $k\le\lg(n)$ times so that we have a polynomial
$h$ of degree $n$.  Once we obtain $h$, we can reduce this polynomial
directly, by writing $h(x)=x^nh' + h''(x)$, where $h'$ is constant and
$h''$ has degree $n-1$, and then adding $h'$ times $x^n$ mod $q$ to
$h''$.

The above discussion for a polynomial $h$ of degree $n-1+2^k$ holds in
particular if $h(x) = x^{n-1+2^k}$; it follows by induction that we
can reduce $x^{n+2^k-1}$ for all $k=0,1,2,\ldots,\lg(n)-1$ in time
$n\polylog(d)$.  Finally, we apply the above again to reduce our
arbitrary polynomial $g$.
\end{proof}

We note that we can find a suitable prime $p$ in time $\poly(d)$ by
testing all numbers from $d$ to $\poly(d)$.  This is a preprocessing
step and the time does not count against the claimed measurement time
of $d\polylog(d)$ or claimed decoding time of $m\polylog(d)$ our
algorithm.  (In fact, time $\polylog(d)$ suffices to find a prime.)
We omit details.

From the preceding lemmas, we conclude:
\begin{thm}
There is an implementation of the Chaining Pursuit algorithm that runs
in time $m\polylog(d)$ and requires total
storage $m\log(d)$ numbers bounded by $\poly(d)$ ({\em i.e.}, $O(\log
d)$ bits).
\end{thm}

\section{Signal Approximation with Chaining Pursuit}\label{sec:decode}

Suppose that the original signal $f$ is well-approximated by a signal
with $m$ nonzero entries (spikes).  The goal of the Chaining Pursuit
algorithm is to use a sketch of the signal to obtain a signal
approximation with no more than $m$ spikes.  To do this, the algorithm
first finds an intermediate approximation $g$ with possibly more than
$m$ spikes, then returns $g_m$, the restriction of $g$ to the $m$
positions that maximize the coefficient magnitudes of $g$.
We call the final step of the algorithm the {\em pruning} step.
The algorithm {\em without} the pruning step will be called {\em Chaining
Pursuit Proper}; we focus on that until Section~\ref{sec:pruning}.

The Chaining Pursuit Proper algorithm proceeds in passes.  In each
pass, the algorithm recovers a constant fraction of the remaining
spikes.  Then it sketches the recovered spikes and updates the data
matrix to reflect the residual signal---the difference between the
given signal and the superposition of the recovered spikes.  After
$O(\log m)$ passes, the residual has no significant entries remaining.

The reason for the name ``Chaining Pursuit'' is that this process
decomposes the signal into pieces with supports of geometrically
decreasing sizes.  It resembles an approach in analysis and probability,
also called chaining, that is used to control the size of a function
by decomposing it into pieces with geometrically decreasing sizes.  A
famous example of chaining in probability is to establish bounds on
the expected supremum of an empirical process
\cite{Tal05:Generic-Chaining}.  For an example of chaining in
Theoretical Computer Science, see
\cite{IN05:Nearest-Neighbor}.

\subsection{Overview of Algorithm}

The structure of the Chaining algorithm is similar to other sublinear
approximation methods described in the literature
\cite{GGIMS02:Near-Optimal-Sparse}.  First, the algorithm identifies
spike locations and estimates the spike magnitudes.  Then it encodes
these spikes and subtracts them from the sketch to obtain an implicit
sketch of the residual signal.  These steps are repeated until the
number of spikes is reduced to zero.  The number $a$ that appears in
the statement of the algorithm is a sufficiently large constant that
will be discussed further in Section \ref{sec:chaining-analysis} and
the quantity $m_k$ is $m/a^k$.  Pseudocode is given in
Figure~\ref{fig:algo}.

\begin{figure}
\caption{Chaining Pursuit algorithm}\label{fig:algo}
\begin{center}
\tt
\begin{tabular}{|p{.9\textwidth}|}
\hline
\centerline{\textbf{Algorithm: Chaining Pursuit}} \\

Inputs: Number $m$ of spikes, the sketch $V$, the isolation matrix $\bm{A}$ \\
Output: A list of $m$ spike locations and values \\
\\
  For each pass $k = 0, 1, \dots, \log_a m$: \\
\hspace{2pc} For each trial $t = 1, 2, \dots, O( k \log d )$: \\
\hspace{4pc} For each measurement $n = 1, \dots, O(m / 2^k)$ \\
\hspace{6pc} Use bit tests to identify the spike position \\
\hspace{6pc} Use a bit test to estimate the spike magnitude \\
\hspace{4pc} Retain $m_k$ distinct spikes with values largest in  magnitude \\
\hspace{2pc} Retain spike positions that appear in more than 9/10 of trials \\
\hspace{2pc} Estimate final spike sizes using medians \\
\hspace{2pc} Encode the spikes using the measurement operator \\
\hspace{2pc} Subtract the encoded spikes from the sketch \\
  Return the signal consisting of the $m$ largest retained spikes.\\
\hline
\end{tabular}
\vspace{1pc}
\end{center}
\end{figure}

\subsection{Implementation}

Most of the steps in this algorithm are straightforward to implement
using standard abstract data structures.  The only point that requires
comment is the application of bit tests to identify spike positions
and values.

Recall that a measurement is a row of the sketch matrix, which
consists of $\log_2 \lceil d \rceil + 1$ numbers:
$$
\left[\begin{array}{cccc|c}
b(0) & b(1) & \dots & b( \log_2 \lceil d \rceil - 1 ) & c
\end{array}\right].
$$
The number $c$ arises from the top row of the bit test matrix.
We obtain an (estimated) spike location from these numbers as follows.
If $\abs{b(i)} \geq \abs{c - b(i)}$, then the $i$th bit of the
location is zero.  Otherwise, the $i$th bit of the location is one.
To estimate the value of the spike from the measurements, we use $c$.

Recall that each measurement arises by applying the bit test matrix to
a copy of the signal restricted to a subset of its components.  It is
immediate that the estimated
location and value are accurate if the subset contains a single large
component of the signal and the other components have smaller $\ell_1$
norm.

We encode the recovered spikes by accessing the columns of the
isolation matrix corresponding to the locations of these spikes and
then performing a sparse matrix-vector multiplication.  Note that
this step requires random access to the isolation matrix.

\subsection{Storage costs}

The primary storage cost derives from the isolation matrix $\mtx{A}$.
Otherwise, the algorithm requires only $O(m \log d)$ working space.

\subsection{Time costs}

During pass $k$, the primary cost of the algorithm occurs when we
encode the recovered spikes.  The number of recovered spikes is at
most $O(m / a^k)$, so the cost of encoding these spikes is $O(m a^{-k}
\log^2 (m) \log^2 (d))$.  The cost of updating the sketch is the same.
Summing over all passes, we obtain $O( m \log^2 (m) \log^2(d) )$ total
running time.

\section{Analysis of Chaining Pursuit} \label{sec:chaining-analysis}

This section contains a detailed analysis of the Chaining Pursuit
Proper algorithm ({\em i.e.}, Chaining Pursuit without the final
pruning step), which yields the following theorem.  Fix an isolation
matrix $\mtx{A}$ which satisfies the conclusions of
Condition~\ref{cond:isolation-matrix} in the sequel and let
$\Fee=\mtx{A}\rtp\mtx{B}$, where $\mtx{B}$ is a bit test matrix.

\begin{thm}[Chaining Pursuit Proper] Suppose that $f$ is a $d$-dimensional
signal whose best $m$-term approximation with respect to $\ell_1$ norm
is $f_m$.  Given the sketch $V = \Fee f$ and the matrix $\Fee$,
Chaining Pursuit Proper produces a signal $\widehat{f}$ with at most $O(m)$
nonzero entries.  This signal estimate satisfies
\[
\smnorm{1}{ f - \widehat{f} }
	\leq (1 + C \log m) \smnorm{1}{ f - f_m }.
\]
In particular, if $f_m = f$, then also $\widehat{f} = f$.
\label{thm:chaining}
\end{thm}

\subsection{Overview of the analysis}

Chaining Pursuit Proper is an iterative algorithm.  Intuitively, at
some iteration $k$, we have a signal that consists of a limited number
of spikes (positions whose coefficient is large) and noise (the
remainder of the signal).  We regard 
the application of the isolation matrix $\mtx{A}$ as repeated trials
of partitioning the $d$ signal positions into $\Theta(m_k)$ random
subsets, where $m_k$ is approximately the number of spikes, and
approximately the ratio of the 1-norm of the noise to the magnitude of
spikes.  There are two important phenomena:
\begin{itemize}
\item A measurement may have exactly one spike, which we call
  \term{isolated}.
\item A measurement may get approximately its fair share of the
  noise---approximately the fraction $1/\mu$ if $\mu$ is the number of
  measurements.
\end{itemize}
If both occur in a measurement, then it is easy to see that the bit
tests will allow us to recover the position of the spike and a
reasonable estimate of the coefficient (that turns out to be accurate
enough for our purposes).  With high probability, this happens to many
measurements.

Unfortunately, a measurement may get zero spikes, more than one spike,
and/or too much noise.  In that case, the bit tests may return a
location that does not correspond to a spike and our estimate of the
coefficient may have error too large to be useful.  In that case, when
we subtract the ``recovered'' spike from the signal, we actually
introduce additional spikes and \term{internal} noise into the
signal.  We bound both of these phenomena.  If we introduce a
false spike, our algorithm has a chance to recover it in future
iterations.   If we introduce a false position with small magnitude,
however, our algorithm may not recover it later.  Thus the internal
noise may accumulate and ultimately limit the performance of our
algorithm---this is the ultimate source of the logarithmic factor in
our accuracy guarantee.

In pass $k = 0$, the algorithm is working with measurements of the
original signal $f$.  This signal can be decomposed as $f = f_m + w$,
where $f_m$ is the best $m$-term approximation of $f$ ({\em spikes})
and $w$ is the remainder of the signal, called \term{external noise}.
If $w = 0$, the analysis becomes quite simple. Indeed, in that case we
exactly recover a constant fraction of spikes in each pass; so we will
exactly recover the signal $f$ in $O(\log m)$ passes.  In this
respect, Chaining is superficially similar to, {\em
  e.g.},~\cite{GGIMS02:Near-Optimal-Sparse}.  An important difference
is that, in the analysis of Chaining pursuit, we exploit the fact that
a fraction of spikes is recovered except with probability
exponentially small in the number of spikes; this lets us unite over
all configurations of spike positions and, ultimately, to get a
uniform failure guarantee.

The major difficulty of the analysis here concerns controlling the
approximation error from blowing up in a geometric progression from
pass to pass.  More precisely, while it is comparatively easier to
show that, for {\em each} signal, the error remains under control,
providing a uniform guarantee---such as we need---is more challenging.
In presence of the external noise $w \ne 0$, we can still recover a
constant fraction of spikes in the first pass, although with error
whose $\ell_1$ norm is proportional to the $\ell_1$ norm of the noise
$w$.  This error forms the ``internal noise'', which will add to the
external noise in the next round. So, {\em the total noise doubles at
every round}.  After the $\log_a m$ rounds (needed to recover all
spikes), the error of recovery will become polynomial in $m$.  This is
clearly unacceptable: Theorem \ref{thm:chaining} claims the error to
be logarithmic in $m$.

This calls for a more delicate analysis of the error.  Instead of
adding the internal noise as a whole to the original noise, we will
show that the internal noise spreads out over the subsets of the
random partitions.  So, most of the measurements will contain a small
fraction of the internal noise, which will yield a small error of
recovery in the current round.  The major difficulty is to prove that
this spreading phenomenon is {\em uniform}---one isolation matrix
spreads the internal noise for all signals $f$ at once, with high
probability.  This is a quite delicate problem.  Indeed, in the last
passes a constant number of spikes remain in the signal, and we have
to find them correctly. So, the spreading phenomenon must hold for all
but a constant number of measurements.  Allowing so few exceptional
measurements would naturally involve a very weak probability of such
phenomenon to hold. On the other hand, in the last passes the internal
noise is very big (having accumulated in all previous passes). Yet we
need the spreading phenomenon to be uniform in all possible choices of
the internal noise.  It may seem that the weak probability estimates
would not be sufficient to control a big internal noise in the last
passes.

We will resolve this difficulty by doing ``surgery'' on the internal
noise, decomposing it in pieces corresonding to the previous passes,
proving corresponding uniform probability estimates for each of these
pieces, and uniting them in the end.  This leads to
Condition~\ref{cond:isolation-matrix}, which summarizes the needed properties
of the
isolation matrix.

The proof of Theorem \ref{thm:chaining} is by induction on the pass
$k$.  We will normalize the signal so that $\pnorm{1}{w} =
1/(400000a)$.  We will actually prove a result stronger than Theorem
\ref{thm:chaining}.  The following is our central loop invariant:

\begin{inv}\label{inv:inductive-hypoth} 
In pass $k$, the signal has
the form
\begin{equation}\label{eqn:inductive-hypoth} 
f^{(k)} = s_k + w + \sum_{j =0}^{k-1} \nu_j
\end{equation}
where $s_k$ contains at most $m_k$ spikes, $w=f-f_m$ is the external
noise, and each vector $\nu_j$ is the
\term{internal noise} from pass $j$, which consists of $3 m_j$ or
fewer nonzero components with magnitudes at most $2/m_j$.
\end{inv}
When we
have finished with all passes (that is when $k = 1+ \log_a m$), we will
have no more spikes in the signal ($m_k = 0$ thus $s_k = 0$). This at
once implies Theorem \ref{thm:chaining}.

The proof that Invariant~\ref{inv:inductive-hypoth} is maintained will
only use two properties of an isolation matrix, given in
Condition~\ref{cond:isolation-matrix}.  While we only know how to
construct such matrices using randomness, any matrix satisfying these
properties is acceptable.  Section~\ref{sec:determ} will prove that
Invariant~\ref{inv:inductive-hypoth} holds for any matrix $\Fee$
having the properties in Condition~\ref{cond:isolation-matrix};
Section~\ref{sec:prob} proves that most matrices (according to the
definition implicit in Section~\ref{sec:isolation}) satisfy these
properties.  Note that the conditions are given in terms of matrix
actions upon certain kinds of signals, but the conditions are
properties only of matrices.

\begin{cond}[Chaining Recovery Conditions for Isolation Matrices]
  \label{cond:isolation-matrix}
  A 0-1 matrix with pass/trial hierarchical structure described
  in Section~\ref{sec:isolation} ({\em i.e.}, any matrix from the sample
  space described in Section~\ref{sec:isolation})
  is said to satisfy the \term{Chaining Recovery Conditions} if
  for any signal of the form in
  Invariant~\ref{inv:inductive-hypoth} and for any pass $k$, then at least 99/100
  of the trial submatrices have these two properties:
\begin{enumerate} 
\item All but $\tfrac{1}{100} m_{k+1}$ spikes appear alone in a
  measurement, isolated from the other spikes.
\item Except for at most $\tfrac{1}{100} m_{k+1}$ of the measurements,
the internal and external noise assigned to each measurement has
$\ell_1$ norm at most $\tfrac{1}{1000} m_k^{-1}$.  
\end{enumerate}
\end{cond}

\subsection{Deterministic Part}~\label{sec:determ}

In this section, we consider only matrices satisfying
Condition~\ref{cond:isolation-matrix}.
Proposition~\ref{prop:no-exceptions} considers the performance of the
algorithm in one of the 99/100 non-exceptional trials under an
artificial assumption that will be removed in
Proposition~\ref{prop:one-trial}.  Following that, we consider the
performance of the combination of trials, prove that
Invariant~\ref{inv:inductive-hypoth} is maintained, and conclude about
the overall performance of Chaining Pursuit Proper.

\begin{prop}[One Trial, No Inaccuracies] \label{prop:no-exceptions}
Suppose that a trial is not exceptional.  Assume that each
measurement contains at most one spike and that the external noise in
each measurement is no greater than $\eps = \tfrac{1}{1000} m_k^{-1}$.
Then the trial constructs a list of at most $m_k$ spikes.
\begin{enumerate}
\item If $\abs{f^{(k)}(i)} > 2\eps$ then the list
contains a spike with position $i$ and estimated value $f^{(k)}(i) \pm
\eps$.
\item If the list contains a spike with position $i$ and
$\abs{f^{(k)}(i)} \leq 4\eps$, then the estimated value of the spike
is no more than $5\eps$ in magnitude.
\end{enumerate}
We call list items that satisfy these estimates \term{accurate}.
\end{prop}
\begin{proof}
To prove this proposition, we outline a series of lemmas.  We begin
with a simple observation about the performance of the bit-tests.
\begin{lemma}
Assume that a measurement contains a position $i$ of value $p$ and
that the $\ell_1$ norm of the other positions in the measurement is at
most $\epsilon$.  Then
\begin{enumerate}
\item The estimated value $\pest$ is bounded by the total measurement;
that is, $|\pest| \leq |p| + \epsilon$.
\item If $|p| > 2 \epsilon$, then the estimated position is $i$ ({\em i.e.},
the bit-test locates the position correctly) and the estimated value
is within $\epsilon$ from $p$, $|\pest - p| \leq \epsilon$.
\end{enumerate}
\label{lemma:simple-bit-test}
\end{lemma}
\begin{proof}
Follows immediately from the definitions of the bit-tests and the
estimation procedures.
\end{proof}

Let us set $\epsilon = \frac{1}{1000} m_k^{-1}$ and observe that
$\frac{1}{m_{k-1}} < \frac{1}{1000 m_k} = \epsilon$.
Definition~\ref{defn:goodmeas} establishes two criteria that
most measurements satisfy.  We refer to these two types as
\term{good measurements} and define them precisely.
\begin{defn}\label{defn:goodmeas}
A
\term{good measurement} satisfies one of the following two criteria:
\begin{enumerate}
\item The measurement is empty; that is, it contains
positions with values $\abs{f^{(k)}(i)} \leq \epsilon$ and the total
$\ell_1$ norm of the positions in the measurement is less than
$1.5 \epsilon$.
\item The measurement contains one spike at position $i$ with
$\abs{f^{(k)}(i)} > \epsilon$ and the $\ell_1$ norm of all other
positions in this measurement is less than $0.5\epsilon$.
\end{enumerate}
\end{defn}

The next lemma states that for good measurements, the bit-tests return
reasonably accurate estimates.
\begin{lemma}
Assume that a measurement is a good one.  If the measurement is empty,
then the estimated value of $f^{(k)}$ in that measurement is no more
than $1.5 \epsilon$.  If the measurement contains one spike, then the
estimated position is the position of the spike and its estimated
value is within $0.5 \epsilon$ of the true value of the spike.
\end{lemma}
\begin{proof}
Follows from the definitions of good measurements and
Lemma~\ref{lemma:simple-bit-test}.
\end{proof}

The next lemma follows from the previous argument and demonstrates
that if the bit-tests identify a spike position, they do so reasonably
accurately and precisely.
\begin{lemma}
Assume that in a single measurement the bit-tests identify and
estimate a spike at position $i$ and that the estimated value $\pest$
is greater than $1.5 \epsilon$, $\abs{\pest} > 1.5 \epsilon$.  Then
the measurement contains a spike at position $i$ and the true value of
$p$ is within $0.5 \epsilon$ of $\pest$.
\end{lemma}

Strictly speaking, we perform multiple trials at each round $k$.  We
obtain, in a single trial, an estimate position and its estimated
value, which we call the preliminary estimated value for that
position.  If, after performing all the trials, we have more than one
preliminary estimated value for an estimated position, we simply use
the preliminary value with the largest absolute value as the estimate
assigned to this position.  We then identify the $m_k$ positions with
the largest assigned estimates.  A simple argument (which we omit here
for brevity) demonstrates that the true value $p$ of a spike at
position $i$ with $\abs{p} > 2 \epsilon$ is assigned an estimate
$\pest$ within $0.5\epsilon$ of $p$.  Furthermore, if the assigned
estimate $\pest$ of a spike at position $i$ satisfies $\abs{\pest} >
1.5\epsilon$, then the true value $p$ is within $0.5\epsilon$ of
$\pest$.  To simplify our arguments in what follows, we simply refer
to the estimated values in one trial as the assigned estimate values,
$\widetilde{f}^{(k)}(i)$.

With the above lemmas, we are able to complete the proof of the
proposition.  Our previous discussion shows that those positions $i$
with $\abs{f^{(k)}(i)} > \epsilon$ include the positions with
estimated values larger than $1.5\epsilon$; {\em i.e.},
\[  \Big\{ i \,\Big|\, |\widetilde{f}^{(k)}(i)| > 1.5 \epsilon \Big\} \subseteq
         \Big\{ i \,\Big|\, |f^{(k)}(i)| > \epsilon \Big\}.
\]
Our inductive hypothesis assumes that there are at most $m_k$
positions in the right set above, so there are at most $m_k$
positions in the left set as well.  Our algorithm (for one trial)
identifies all of these positions and reports estimated values that
are within $0.5\epsilon$ of the true values.  Hence, our list of
identified positions includes those $i$ with $\abs{f^{(k)}(i)} >
2\epsilon$.  If a position $i$ is identified and if $\abs{f^{(k)}(i)}
\leq 4 \epsilon$, the its estimated value is at most $5 \epsilon$ in
magnitude by the previous lemmas.  This proves the proposition.
\end{proof}

The next proposition removes the artificial assumption on the spikes
and noise.
\begin{prop}[One Trial] \label{prop:one-trial} Suppose that the trial
  is not an exceptional trial.  In this trial, suppose each
  measurement is a good one, except for at most $\tfrac{1}{50}
  m_{k+1}$.  Then the trial constructs a list of at most $m_k$ spikes.
  All items in the list are accurate, except at most $\tfrac{3}{50}
  m_{k+1}$.
\end{prop}

\begin{proof}  We begin the proof with a lemma that shows the list
  produced by the algorithm is stable with respect to changes in a few
  measurements.  
\begin{lemma}
  Assume that we perform one trial of the algorithm with two different
  signals and that their measurements (in this one trial) are
  identical except for $b$ measurements.  Then the estimated signals
  are identical except in $2b$ positions.
\end{lemma}  
\begin{proof}[Proof sketch.]
Prove this for $b = 2$ and then proceed by induction.
\end{proof}

Let us now consider the set of fewer than $\tfrac{1}{50} m_{k+1}$ bad
measurements and set to zero the signal positions that fall into these
measurements.  This procedure creates two signals: the original signal
and the restricted signal (with zeroed out positions).  The restricted
signal satisfies the conditions in
Proposition~\ref{prop:no-exceptions} so all its measurements are good
ones and agree with those of the original signal except for
$\tfrac{1}{50} m_{k+1}$ measurements.  The previous lemma guarantees
that the estimated signals for the original and restricted signals are
identical in all but $\tfrac{1}{25} m_{k+1}$ positions.  By our
inductive hypothesis, there are at most $\tfrac{1}{50} m_{k+1}$
positions of the original signal with value greater than $2 \epsilon$
in the exceptional measurements.  Let us gather these $\tfrac{1}{25}
m_{k+1}$ and $\tfrac{1}{50} m_{k+1}$ exceptional positions into one
set of $\tfrac{3}{50} m_{k+1}$ exceptions.  It is straightforward to
show that the positions not in this exceptional set are good positions
and, if they are identified, they are identified accurately.
\end{proof}

We combine results from all trials.  The algorithm considers positions
identified in at least $\tfrac{9}{10}$ of the total trials $T$.  It
then takes the median (over all trials) to estimate the values of
these positions.

\begin{lemma}[Combining Trials] \label{lem:combining}
  The number of list items that are inaccurate in more than 1/10 of
  the trials is at most $m_{k+1}$.  The total number of positions that
  appear in 9/10 of the trials is at most $\tfrac{10}{9} m_k$.
\end{lemma}
\begin{proof}
We prove the first part of the lemma with a simple counting argument.
Let $T$ denote the total number of trials.  We have to bound $b$ where
\[  b = \#\{ \text{positions bad in} \geq \tfrac{T}{10} \text{ trials}\}
      \leq \#\{ \text{positions bad in} \geq \tfrac{T}{11} \text{ good
        trials}\}.
\]
Let 
\[  \Sigma = \sum_{j \in \text{ good trials}} \#\{\text{positions bad
  in trial $j$}\}.
\]
We have $\Sigma \geq \tfrac{b T}{11}$.  Let $T'$ be the number of good
trials.  Then 
Proposition~\ref{prop:one-trial} tells us that $\Sigma \leq
\tfrac{3}{50} m_{k+1} T' \leq \tfrac{3 \cdot T}{50} m_{k+1}$.
Therefore, $\tfrac{b T}{11} \leq \tfrac{3 T}{50} m_{k+1}$ and, hence, $b
\leq \tfrac{33}{50} m_{k+1}$.

To prove the second part, recall that the algorithm updates a position
if and only if the position is identified in at least 9/10 of the
trials.  Let $\ell$ denote the number of such positions.  In every
trial, $m_k$ positions are identified.  Hence,
\[ m_k T = \sum_{t = 1}^T \#\{ \text{positions identified in trial $t$}\}
         \geq \frac{9}{10} T \ell
\]
and thus $\ell \leq \tfrac{10}{9} m_k$.
\end{proof}

Now we are ready to prove the induction step.  Recall that after round
$k$, the new signal is the difference between the current signal and
its estimate: $f^{(k)} = f^{(k-1)} - \tilde f^{(k-1)}$, with the
convention that if a signal position is not considered and not changed
by the algorithm, its estimated value is zero.

\begin{lemma}[Induction Hypothesis] \label{lem:ind-hyp} 
  After pass $k$, there are at most $m_{k+1}$ spikes remaining.  The
  contribution $\nu_k$ to the internal noise contains at most $3m_k$
  components with values at most $2/m_k$.
\end{lemma}
\begin{proof}
Recall that $4.5 \epsilon < m^{-1}_k$.  It suffices to prove that for
the non-exceptional positions $i$ that satisfy the conclusions of
Lemma~\ref{lem:combining}, the value $\abs{f^{(k)}(i)} \leq 4.5
\epsilon <
\tfrac{2}{m_k}$.  Let us fix such a position $i$ which is good in at
least 9/10$T$ trials and show that 
\[   \abs{f^{(k-1)}(i) - \tilde f^{(k-1)}(i)} \leq 4.5 \epsilon.
\]

If $\abs{f^{(k-1)}(i)} > 2\epsilon$, then the goodness of $i$ in 9/10 trials
implies that $i$ is identified in these trials and hence $i$ is
considered by the Algorithm.  In each of these $9/10$ trials, the
goodness of $i$ also means that the assigned estimated value of $i$ is
within $0.5\epsilon$ from its true value $f^{(k-1)}(i)$.  Since
$\tilde f^{(k-1)}(i)$ is the median of the assigned estimated values
of $i$ in each trial, it follows that it is also within $0.5\epsilon$
from the true value $f^{(k-1)}(i)$.

Suppose that $\abs{f^{(k-1)}(i)} \leq 2\epsilon$.  If $i$ is not
considered by the algorithm, then the value of the signal at this
position is not changed, so $|f^{(k)}(i)| = |f^{(k-1)}(i)| \leq
2\epsilon$.  We can assume that $i$ is considered by the algorithm.
There are $(9/10)T$ trials in which $i$ is identified and there are
$(9/10)T$ trials in which $i$ is good.  Hence, there are at least
$8/10 T$ trials in which $i$ is both good and identified.  By the
definition of goodness, this means that in each of these trials, the
assigned estimated value $|\tilde f^{(k-1)}(i)|$ is at most
$2.5\epsilon$.
Since the estimated value is the median of the assigned
estimated values of $i$ in each trial, it follows that $|\tilde
f^{(k-1)}(i)| \le 2.5\epsilon$. Then
\[  |f^{(k-1)}(i) - \tilde f^{(k-1)}(i)| \leq
    |f^{(k-1)}(i)| + |\tilde f^{(k-1)}(i)| \leq 2\epsilon + 2.5\epsilon
  = 4.5\epsilon.
\]
This completes the proof of the first part of the lemma.  

The first part of this lemma shows formally that the difference
between the spikes in the signal $f^{(k)}$ and the large entries in
the update signal ({\em i.e.}, those with absolute values greater than
$m_k^{-1}$) contains at most $m_{k+1}$ terms.  By the inductive step,
the same holds for the previous rounds.  In addition,
Lemma~\ref{lem:combining} tells us that the algorithm updates at most
$\tfrac{10}{9} m_k$ positions in the signal.  By the triangle inequality it
follows that the difference contains at most $m_{k+1} + m_k + \tfrac{10}{9}m_k
\leq 3m_k$ terms.  The maximal absolute value of the difference signal
is
\[  \frac{1}{m_k} + \frac{1}{m_{k-1}} \le \frac{2}{m_k}.
\]
\end{proof}
The previous lemma proves the induction hypothesis.  
The next and final lemma of this section controls the recovery error
and completes the proof of
Theorem~\ref{thm:isolation-matrix}.
\begin{lemma}[Total Spikes and Recovery Error] \label{lem:rec_error}
Chaining Pursuit Proper recovers at most $O(m)$ spikes.  The total recovery
error is at most $(1 + C \log m) \pnorm{1}{w}$.
\end{lemma}
\begin{proof}[Proof Sketch.]  After pass $K = \log_a m$, there are no
more spikes remaining since $m_k = m/a^k < 1$.  At most $\tfrac{10}{9} m_k$
spikes are recovered in pass $k$.  Since $m_k$ decays geometrically,
the total number of spikes is $O(m)$.  The error after the last pass
is the $\ell_1$ norm of the signal $f^{(K+1)}$.  This signal consists
of the external noise, which has norm $\pnorm{1}{w}$, and the
internal noise, which satisfies $$ \pnorm{1}{ \sum\nolimits_{j=0}^{K}
\nu_j } \leq \sum\nolimits_{j=1}^{K} 6 m_j m_j^{-1} = 6 \log_a m.  $$
Since $a$ is a constant and $\pnorm{1}{w}$ was normalized to be
constant, the overall error is at most $(1+C\log m)\pnorm{1}{w}$ for
some constant $C$.
\end{proof}

\subsection{Probabilistic Part} \label{sec:prob}

Here we prove that a random isolation matrix $\mtx{A}$ indeed
satisfies the CRC with high probability.

\begin{thm}\label{thm:isolation-matrix}
  With probability at least $(1 - O(d^{-3}))$, a matrix $\mtx{A}$ drawn from the
  distribution described in Section~\ref{sec:isolation} satisfies the
  Chaining Recovery Conditions (Conditions~\ref{cond:isolation-matrix}).
\end{thm}

The main lemmas of this section are as follows.  First,
Lemma~\ref{lem:ball-buckets} is an abstract technical Lemma about
putting balls into buckets and the number of isolated balls that
likely result.  Lemma~\ref{lem:isolations} is a corollary for our
context.  We then show, in Lemma~\ref{lem:internal-noise}, that
Condition~\ref{cond:isolation-matrix} holds for most matrices.

\begin{lemma}[Balls and Bins] \label{lem:ball-buckets}
Put $n$ balls randomly and independently into $N > C(M) n$ buckets.
  Then, with probability $1 - 2e^{-9n}$,
  all except $n/M$ balls are isolated in their buckets.
\end{lemma}
\begin{proof} 
  One complication in the proof comes from the absence of independence
  among buckets.  We would rather let buckets choose balls. However,
  the contents of different buckets is dependent because the total
  number of balls is limited.  So we will replace the original
  $n$-ball model
  with an {\em independent} model.  The independent
  model will be easier to handle by the standard large deviation
  technique; the independent model reduces the original model from it
  by conditioning on the number of balls.

  The independent model is the following assignment. We divide each
  bucket into $n$ sub-buckets, and let $\delta_{ki}$ be independent
  $0,1$ valued random variables with expectation $\E \delta_{ki} =
  1/N$, for all buckets $k=1,\ldots,N$ and sub-buckets $i=1,\ldots,n$.
  The independent random variables $X_k = \sum_{i=1}^n \delta_{ki}$
  will be called the number of balls in bucket $k$ in the independent
  model.  If we condition on the total number of such ``balls'', we
  obtain the distribution of the numbers of true balls $X'_k$ in
  bucket $k$ in the original model:
\[  (X'_1, \ldots, X'_N) \equiv \Big( X_1, \ldots, X_N \; \Big| \;
            \sum_{k=1}^N X_k = n \Big).
\]

To prove the Lemma, we have to show that the number of non-isolated
balls is small. The number of non-isolated balls in bucket $k$ is $
Y'_k = X'_k \cdot 1_{\{X'_k > 1\}}$ so the conclusion of the Lemma is
that
\begin{equation}\label{eqn:non-isolated original}
  \Prob \Big\{ \sum_{k=1}^N Y'_k > n/M \Big\}
  \le 2 e^{-9n}.
\end{equation}

We will now transfer this problem to the independent model.
First, without loss of generality we can change the $n$ balls in
the lemma and in the original model to $0.9 n$ balls.
We do not change the independent model, so the number of
non-isolated balls in the independent model is $Y_k = X_k \cdot
1_{\{X_k > 1\}}$.  We have to bound
\begin{align*}
\Prob \Big\{ \sum_{k=1}^N Y'_k > n/M \Big\} 
   &= \Prob \Big\{ \sum_{k=1}^N Y_k > n/M
       \Big|  \sum_{k=1}^N X_k = 0.9 n \Big\} \\
   &\le \Prob \Big\{ \sum_{k=1}^N Y_k > n/M
       \Big| \sum_{k=1}^N X_k \ge 0.9 n \Big\} \\
   &\le \frac{\Prob \Big\{ \sum_{k=1}^N Y_k > n/M \Big\} }
      {\Prob \Big\{ \sum_{k=1}^N X_k \ge 0.9 n \Big\}}.
\end{align*}
By Prokhorov-Bennett inequality,
\[  \Prob \Big\{ \sum_{k=1}^N X_k \ge 0.9 n \Big\}
      = \Prob \Big\{ \sum_{k=1}^N \sum_{i=1}^n \delta_{ki} \ge 0.9 n \Big\}
\ge 1/2.
\]
Therefore,
proving Equation~\eqref{eqn:non-isolated original} in the
original model reduces to proving that
\begin{equation} \label{eqn:non-isolated independent}
  \Prob \Big\{ \sum_{k=1}^N Y_k > n/M \Big\} \le e^{-9n}
\end{equation}
in the independent model.

A standard way to prove deviation inequalities such as
Equation~\eqref{eqn:non-isolated independent} is through the moment
generating function. By Markov's inequality and independence of the
variables $Y_k$, we have
\begin{align*}
\Prob \Big\{ \sum_{k=1}^N Y_k > n/M \Big\}
  &= \Prob \Big\{ e^{10 M \sum_{k=1}^N Y_k} > e^{10 n} \Big\} 
  \le e^{-10 n} \cdot \E \Big[ e^{10 M \sum_{k=1}^N Y_k} \Big] \\
  &= e^{-10 n} \cdot \Big(\E \Big[ e^{10 M Y_1}\Big] \Big)^N
\end{align*}
To complete the proof of Equation~\eqref{eqn:non-isolated independent}
it remains to show that for $Y = ( \sum_{i=1}^n \delta_i ) \cdot
1_{\{\sum_{i=1}^n \delta_i > 1\}}$, its moment generating function satisfies
\begin{equation}                \label{eqn:moment generating}
  \Big(\E \Big[e^{10 M Y}\Big]\Big)^N \le e^n
\end{equation}
where $\delta_i$ are $0,1$ valued independent random variables with
$\E \delta_i = 1/N$.  To estimate the moment generating function $\E
[e^{MY}]$ in Equation~\eqref{eqn:moment generating}, it suffices to
know the tail probability $\Prob \{ Y > t \}$ for large $t$.  For
large $t$, we estimate this tail probability by removing the
restriction onto non-isolated bucket in the definition of $Y$ and
applying Chernoff's inequality for independent random variables.  The
tail probability is, however, much smaller if we do restrict onto
non-isolated buckets. We take this into account for small $t$ by
computing the expectation of $Y$ (which is straightforward).

Let us start with the first moment of $Y$.  We claim that
\begin{equation} \label{eqn:first moment}
  \E [Y] \le C \Big( \frac{n}{N} \Big)^2.
\end{equation}
Compare this with the average number of balls without conditioning on
being non-isolated, $\E [X_k] = \frac{n}{N}$.  Indeed, by the
linearity of expectation,
\begin{align*}
  \E [Y]
  &= n \cdot \E \Big[\delta_1 \cdot 1_{\{\sum_{i=1}^n \delta_i > 1\}}\Big] \\
  &= n \cdot \Prob \Big \{ \delta_1 = 1
  \ \text{and there exists} \ i \in \{2, \ldots, n\} : 
    \ \delta_i = 1 \Big\} \\
  &= \Prob \Big\{ \delta_1 = 1 \Big\}
  \cdot \big( 1 - 
      \Prob \Big\{ \forall i \in \{2, \ldots, n\} :\delta_i = 0 \Big\} \big) \\
  &= \frac{n}{N} \cdot \Bigg( 1 - \Big( 1 - \frac{1}{N} \Big)^{n-1}
  \Bigg) 
  \le  \frac{n}{N} \cdot \Big(1 - e^{-n/N}\Big) 
  \le C \Big( \frac{n}{N} \Big)^2.
\end{align*}
Next, by the Chernoff inequality, for $s > 2$, we have
\begin{equation}   \label{eqn:tail}
  \Prob \{ Y > s \}
  \le \Prob \{ \sum_{i=1}^n \delta_i > s \}
  \le ( s \frac{N}{n} )^{-s}.
\end{equation}
Now we are ready to bound the moment generating function.
Let $K = 10 M$ and change variables $t = e^{Ks}$, so that
\begin{align*}
\E \Big[e^{KY}\Big]
  & = \int_0^\infty \Prob \Big\{ e^{KY} > t \Big\} \; dt 
  = 1 + \int_1^\infty \Prob \Big\{ e^{KY} > t \Big\} \; dt \\
  &= 1 + K \int_0^\infty \Prob \Big\{ Y > s \Big\} \; e^{Ks} \; ds.
\end{align*}
We split the integral in two parts.
We use Equation~\eqref{eqn:first moment} to estimate the integral
near zero as
\begin{align*}
\int_0^3 \Prob \Big\{ Y > s \Big\} \; e^{Ks} \; ds
  &\le e^{3K} \int_0^\infty \Prob \Big\{ Y > s \Big\} \; ds \\
  &= e^{3K} \E[Y]
  \le C e^{3K} \Big( \frac{n}{N} \Big)^2,
\end{align*}
and we use Equation~\eqref{eqn:tail} to estimate the integral near infinity as
\begin{align*}
\int_3^\infty \Prob \Big\{ Y > s \Big\} \; e^{Ks} \; ds
  &\le \int_3^\infty \Big( \frac{N}{n} s\Big)^{-s} e^{Ks} \; ds 
  \le \int_3^\infty \Big(  \frac{N}{n} e^{-K}\Big)^{-s} \; ds \\
  &= \frac{1}{\ln ( \frac{N}{n}  e^{-K})}
    \Big(  \frac{N}{n} e^{-K}\Big)^{-3} 
  \le e^{2K} \Big( \frac{n}{N} \Big)^2.
\end{align*}
Combining these, we conclude that
\[ \E \Big[e^{KY}\Big]
       \le 1 + C K e^{3K} \Big( \frac{n}{N} \Big)^2
       \le 1 + \frac{n}{10 N}.
\]
Hence we obtain Equation~\eqref{eqn:moment generating}
\[  \Big( \E \Big[e^{KY}\Big]\Big)^N
         \le \Big( 1 + \frac{n}{10 N} \Big)^N
         \le e^{\frac{n}{5N} \cdot N}
         \le e^n.
\]
This proves the lemma.
\end{proof}

\begin{lemma}[Isolations]\label{lem:isolations}
  Fix a round $k$.  With probability at least $1 - \exp\{-4m_k \log
  d\}$, the following is true.  In pass $k$, at least 99/100 of the
  trial submatrices isolate all but $\tfrac{1}{100}$ of the $m_k$
  spikes.
\end{lemma}

\begin{proof}[Proof sketch.]  
  In the hypothesis of Invariant~\ref{inv:inductive-hypoth}, the
  signal has at most $n := m_k$
  positions of value larger than $1/m_{k-1}$. We put these in $N :=
  \overline{m}_k = C'(a) m /2^k$ buckets.  We can choose the function
  $C'(a)$ so that, for $C$ of Lemma~\ref{lem:rec_error}, we have
  $C'(a) = C(\frac{1}{100 a})$.  Apply Lemma~\ref{lem:ball-buckets}
  which states that all except $\tfrac{1}{100} m_{k+1}$ positions are
  isolated with probability $1 - \delta$, where $\delta = 2e^{-9m_k}$.

Let $I$ be the event that all but $\tfrac{1}{100} m_k$ of the spikes
are isolated.  Let us repeat the random assignment above independently
$T$ times (for $T$ trials) and let $\delta_t$ be independent Bernoulli
random variables, $\E\delta_t = \delta$.  We see by Chernoff's
inequality that
\begin{align*}
  \Prob \{ \text{$I$ fails in more than $\frac{1}{100} T$ trials}
  \} 
  &\le \Prob \Big\{ \sum_{t=1}^T \delta_t > \frac{1}{100} T \Big\} 
  \le (100 e \delta)^{\frac{1}{100} T} \\
  & = \exp \big(-m_k \cdot
  \frac{1}{100} T \big) 
  \le \exp(-4 m_k \log d).
\end{align*}
\end{proof}
This concludes the proof of Lemma~\ref{lem:isolations}.

We now proceed to prove that Invariant~\ref{inv:inductive-hypoth} is
maintained by most isolation matrices $\mtx{A}$.  That is, we need to
show that, for most $\mtx{A}$, when the Chaining Pursuit algorithm
uses $\mtx{A}$ on a 
signal satisfying Invariant~\ref{inv:inductive-hypoth} for round $k$,
the algorithm produces a signal satisfying
Invariant~\ref{inv:inductive-hypoth} for round $k+1$.  So, in the
remainder of this section, we may fix a signal satisfying
Invariant~\ref{inv:inductive-hypoth} for round $k$.
Lemma~\ref{lem:external-noise} controls the external noise and
Lemma~\ref{lem:internal-noise} controls
the internal noise.

\begin{lemma}[External noise] \label{lem:external-noise}
  In pass $k$, in every trial, the number of measurements where the
  $\ell_1$ norm of the external noise exceeds $\tfrac{1}{2000}
  m_k^{-1}$ is at most $\tfrac{1}{200} m_{k+1}$.
\end{lemma}

\begin{proof}[Proof sketch.]  
  This is an easy part of the argument.  The $(1, 1)$ operator norm of
  each matrix $\bm{A}^{(k)}_t$ equals one, so it does not inflate the
  norm of the external noise.  We use Markov's inequality to bound the
  number of measurements with too much noise.  
\end{proof}

\begin{lemma}[Internal noise] \label{lem:internal-noise}
  Fix a round
  $k$.  With probability at least $1 - \exp\{-4m_k \log d\}$, the
  following is true.  In pass $k$, in at least 99/100 of the trials, the
  number of measurements where the $\ell_1$ norm of the internal noise
  exceeds $\tfrac{1}{2000} m_k^{-1}$ is at most $\tfrac{1}{200}
  m_{k+1}$.
\end{lemma}

\begin{proof}
  Let us recall the Invariant~\ref{inv:inductive-hypoth}.  This lemma
  is a statement is
  about the signal $f^{(k-1)}$ in the positions with values smaller
  than $1/m_{k-1}$.  Lemma~\ref{lem:external-noise} gives us a proof
  for $k = 0$.

  Let $k \ge 1$. We may assume that the external noise $w$ is 0 and we
  may absorb the spikes in
  Equation~\ref{eqn:inductive-hypoth} into
  the first term; {\em i.e.} we can
  assume that our signal has the form
  \begin{equation} \label{eqn:signal} 
          f^{(k)} = \sum_{j=0}^{k-1}  \nu_j
\end{equation}
where $\nu_j$ consists of $4m_j$ or fewer nonzero components with
magnitudes at most $\tfrac{3}{m_j}$.  

To prove this result, we introduce positive parameters $\lambda_j$,
$\epsilon_j$, $j = 0, \ldots, k-1$, which satisfy
\begin{equation}                        \label{eqn:lambda}
  \sum_{j=0}^{k-1} \lambda_j \le \frac{1}{C'}
\end{equation}
and
\begin{equation}                        \label{eqn:epsilon}
  \sum_{j=0}^{k-1} \epsilon_j \le \frac{1}{C'a}
\end{equation}
where $C'$ is a positive absolute constant to be chosen later.
Next, we will prove the following separate claim about the internal noise
$\nu_j$.
\begin{claim}
  Assume that a signal satisfies Equation~\eqref{eqn:signal}.
  Let $j \in \{0, \ldots, k-1\}$. Then
  \begin{equation}                        \label{eqn:round j content}
    \sum_{t=1}^T \#\Big(
      \text{measurements in trial $t$ s.t. $\|\nu_j\|_{1} > 
          \frac{\lambda_j}{m_k}$} \Big)
    < \epsilon_j m_k T
  \end{equation}
  with probability
  \begin{equation}                       \label{eqn:probability}
    1 - e^{-\gamma m_j T},
  \end{equation}
  where $\gamma$ is some positive number such that
  \begin{equation}                        \label{eqn:gamma}
    \gamma T \ge 10 \log d.
  \end{equation}
\end{claim}
\begin{proof}
\noindent {\bf Claim implies Lemma~\ref{lem:internal-noise}.}
We will first show that this Claim implies
Lemma~\ref{lem:internal-noise}.  Assume the claim holds. By the
definition of $T$, the exceptional probability is
$$
e^{-\gamma m_j T}
\le e^{-10 m_j \log d}
\le \binom{d}{4m_j}^{-4},
$$
while the number of choices of round $j$ signal is $\binom{d}{4m_j}$.
Hence, with probability $1 - \binom{d}{4m_j}^{-3}$, the inequality in
Equation~\eqref{eqn:round j content} holds uniformly for all choices of
internal noise $\nu_j$.  Summing up these exceptional probabilities
for all rounds $j = 0, \ldots, k-1$, we conclude that:

\begin{quote}
  {\em with probability $1 - \binom{d}{4m_j}^{-2}$, the system of
    measurements is such that the inequality in
    Equation~\eqref{eqn:round j content} holds uniformly for all
    choices of the signal satisfying Equation~\eqref{eqn:signal}.}
\end{quote}

\noindent Fix a system of measurements which meets these requirements
so that we can dismiss the probability issues.  Let us also fix a
measurement $v$ and a trial $t$.  We refer to $\nu^t_{k,v}$ as the
signal in measurement $v$ and trial $t$.  By
Equation~\eqref{eqn:lambda}, for a fixed trial and a fixed measurement
$v$, we have the containment of events:
\[
  \Big\{ \|\nu^t_{k,v} \|_1 > \frac{1}{2000 m_k} \Big\} 
  \subseteq \bigcup_{j=0}^{k-1}
  \Big\{ \|\nu^t_{j,v} \|_1  > \frac{\lambda_j}{m_k} \Big\}.
\]
Counting the measurements that satisfy each side of this containment,
then summing over the trials, we obtain:
\begin{align*}
   \sum_{t=1}^T \# \Big\{ \|\nu^t_{k,v} \|_1  > \frac{1}{2000 m_k} \Big\} 
   &\le \sum_{t=1}^T \sum_{j=0}^{k-1}
       \# \Big\{  \|\nu^t_{j,v} \|_1  > \frac{\lambda_j}{m_k} \Big\} \\
   &\le \sum_{j=0}^{k-1} \epsilon_j m_k T
  \ \ \ \text{by \eqref{eqn:round j content}} \\
   &\le \frac{1}{C'a} m_k T
  \ \ \ \text{by \eqref{eqn:epsilon}} \\
   &= \frac{1}{C'} m_{k+1} T.
\end{align*}
By Markov's inequality, this implies (provided $C'$ is chosen large enough)
that in at most $\frac{1}{100} T$ trials $t$, the number of
measurements where the $\ell_1$ norm of the internal noise exceeds
$\tfrac{1}{2000 m_k}$ is greater than $\tfrac{1}{200}m_{k+1}$.
This implies the conclusion of Lemma~\ref{lem:internal-noise}.

\noindent{\bf Proof of the claim.}  Now we prove the claim itself.
This is a purely probabilistic problem.  We call the nonzero positions
of $\nu_j$ ``balls'' and we informally refer to the measurements as
``buckets''.  By the definition of $\nu_j$, the $1$-norm of $\nu_j$ in
a measurement (or bucket) $v$ is large, $\| \nu_{j,v} \|_1 >
\frac{\lambda_j}{m_k}$ if and only if the measurement contains at
least $\tfrac{1}{3} \lambda_j \frac{m_j}{m_k}$ balls.  Then
Equation~\eqref{eqn:round j content} is equivalent to
\begin{equation}    \label{eqn:round j balls}
  \sum_{t=1}^T \#(
    \text{measurement in trial $t$ which contain
          $> \frac{1}{3} \lambda_j \frac{m_j}{m_k}$ balls} )
  \le \epsilon_j m_k T.
\end{equation}
To prove this with required probability~\eqref{eqn:probability}, we
will transfer the problem to an independent model---similar to the
proof of Lemma~\ref{lem:ball-buckets}.

Recall that the {\em original model} with $n$ balls and $T$ trials, in
which we want to prove Equation~\eqref{eqn:round j balls}, is to put
$n = 4m_j$ balls into $N = \overline{m}_k$ buckets (or measurements)
independently, and repeat this $T$ times (trials) independently.

We want to replace this by the following {\em independent model},
where the contents of buckets are independent.  There are $N$ buckets
in each of $T$ trials.  Divide each bucket into $S = nT$ sub-buckets.
Let $\delta_{tli}$ be independent $0,1$ valued random variables with
expectation $\E \delta_{tli} = 1/NT$, for all trials $t=1,\ldots,T$,
buckets $l=1,\ldots,N$ and sub-buckets $i=1,\ldots,S$.  The
independent random variables
\begin{equation}                        \label{eqn:Xtl}
  X_{tl} = \sum_{i=1}^S \delta_{tli}
\end{equation}
will be called the number of balls in bucket $l$, trial $t$,
in the independent model.
Note that
\[ \E X_{tl} = \frac{S}{NT} = \frac{n}{N}.
\]
Thus the average total number of balls in buckets in one trial is $n$.
Let $E_{\rm tot}$ be the event that in each of at least $T/2$ trials,
the total number of balls in buckets is at least $n/2$.  It is then
easy to deduce by Chernoff and Prokhorov-Bennett's inequalities that
with probability at least $1/2$ (actually $1 - e^{-nT}$), $E_{\rm
  tot}$ holds; that is, $\Prob \{ E_{\rm tot} \} \ge 1/2$.
We have to bound above the probability of the event
$$
E := \{ \text{ Equation~\eqref{eqn:round j balls} does not hold} \}
$$
in the original model. We reduce it to the independent model as follows:
\begin{align*}
  \Prob \{ \text{E in independent model} \} &\ge \Prob \{ \text{E in
    independent model} \ \big| \ E_{\rm tot} \}
  \cdot \Prob \{ E_{\rm tot} \} \\
  &\ge \frac{1}{2} \; \Prob \{ \text{E in independent model} \ \big| \
      E_{\rm tot} \}
\end{align*}
The probability can only decrease when we consecutively do the
following changes:
\begin{enumerate}
\item remove both occurences of ``at least'' in $E_{\rm tot}$, resulting
  in exactly $T/2$ trials and exactly $n/2$ balls;
\item restrict the sum in Equation~\eqref{eqn:round j balls} to the $T/2$
  trials included in the new (exact) version of $E_{\rm tot}$; and
\item fix the set of $T/2$ trials in $E_{\rm tot}$---say, require that
  these be the first $T/2$ trials.
\end{enumerate}
After doing this, the law becomes the original model with $n/2$ balls and
$T$ trials. Hence,
\[ \Prob \{ \text{E in independent model} \} 
  \ge \frac{1}{2} \;
    \Prob \{ \text{E in original model with $n/2$ balls and $T/2$ trials} \}.
\]
Therefore, it suffices to prove that Equation~\eqref{eqn:round j
  balls} holds in the independent model, with probability as in
\eqref{eqn:probability}, {\em i.e.} with probability $1 - \frac{1}{2}
e^{-\gamma m_j T}$, where $\gamma$ is as in Equation~\eqref{eqn:gamma}.

In order to prove Equation~\eqref{eqn:round j balls} with the
requisite probability, we first estimate the number of balls $X_{tl}$
in one bucket, see Equation~\eqref{eqn:Xtl}.  It is a sum of $S = nT$
independent Bernoulli random variables with expectations
$\frac{1}{NT}$. Then by the Chernoff inequality,
\begin{equation}                        \label{eqn:eta}
  \Prob \Big\{ X_{tl} > \frac{1}{3} \lambda_j \frac{m_j}{m_k} \Big\}
  \le \Big( \frac{1}{12e} \lambda_j \frac{\overline{m}_k}{m_k} \Big)
      ^{- \frac{1}{3} \lambda_j \frac{m_j}{m_k}}.
\end{equation}
Let us call this probability $\eta$.  We have to estimate the sum in
Equation~\eqref{eqn:round j balls} which equals $\sum_{t=1}^T
\sum_{l=1}^N \delta_{tl}$ where 
\[  \delta_{tl} = 1_{\{ X_{tl} >  \frac{1}{3} 
    \lambda_j \frac{m_j}{m_k} \}}
\]
are independent Bernoulli random variables whose expectations are $\E
\delta_{tl} \le \eta $ by Equation~\eqref{eqn:eta}.  Then by the
Chernoff inequality, the probability that Equation~\eqref{eqn:round j
  balls} fails to hold is
\[  \Prob \Big\{ \sum_{t=1}^T \sum_{l=1}^N \delta_{tl} > 
       \epsilon_j m_k T \Big\}
    \le (\beta/e)^{-\epsilon_j m_k T},
\]
where
\begin{equation}                        \label{eqn:beta}
  \beta = \frac{\epsilon_j m_k}{\eta N}
  = \epsilon_j \cdot \frac{m_k}{\overline{m}_k} \cdot \frac{1}{\eta}
  = \epsilon_j
    \cdot \frac{m_k}{\overline{m}_k}
    \cdot \Big( \frac{1}{12e} \lambda_j \frac{\overline{m}_k}{m_k} \Big)
        ^{\frac{1}{3} \lambda_j \frac{m_j}{m_k}}
\end{equation}
To complete the proof, we need to show that
$$
(\beta/e)^{-\epsilon_j m_k T}
\le \frac{1}{2} e^{-\gamma m_j T},
$$
which would follow from
\begin{equation}                       \label{eqn:beta needed}
  (\beta/e)^{\epsilon_j \cdot \frac{m_k}{m_j}}
  \ge e^{2 \gamma},
\end{equation}
where $\gamma$ must satisfy Equation~\eqref{eqn:gamma}.

Now we specify our choice of $\lambda_j$ and $\epsilon_j$. Set
\begin{align*}
\lambda_j &:= \max \Big\{
  12e \Big( \frac{m_k}{\overline{m}_k} \Big)^{1/2}, \;
  24 \frac{m_k}{m_j}, \;
  \frac{1}{C'k}
  \Big\}, \\
\epsilon_j &:= \max \Big\{
  \frac{m_k}{\overline{m}_k}, \;
  \frac{1}{C'ak}
  \Big\}.
\end{align*}
Then Equations~\eqref{eqn:lambda} and \eqref{eqn:epsilon} are clearly
satisfied (recall that $j < k$).  Next, the base in
Equation~\eqref{eqn:beta} is estimated as
$$
\frac{1}{12e} \lambda_j \frac{\overline{m}_k}{m_k}
\ge \Big( \frac{\overline{m}_k}{m_k} \Big)^{1/2}
$$
and the exponent can be estimated using
$$
\lambda_j \frac{m_j}{m_k} \ge 24.
$$
Also, the first factor of Equation~\eqref{eqn:beta} is estimated as
$$
\epsilon_j \cdot \frac{m_k}{\overline{m}_k}
\ge \Big( \frac{m_k}{\overline{m}_k} \Big)^2.
$$
Combining these three estimates, we obtain
$$
\beta
\ge \Big( \frac{m_k}{\overline{m}_k} \Big)^2
  \cdot \Big( \frac{\overline{m}_k}{m_k} \Big)
        ^{\frac{1}{6} \lambda_j \frac{m_j}{m_k}}
\ge \Big( \frac{\overline{m}_k}{m_k} \Big)
        ^{\frac{1}{8} \lambda_j \frac{m_j}{m_k}}.
$$
Now we can check Equation~\eqref{eqn:beta needed}.
\begin{align*}
(\beta/e)^{\epsilon_j \cdot \frac{m_k}{m_j}}
&\ge \Big( \frac{\overline{m}_k}{e m_k} \Big)
        ^{\frac{1}{8} \lambda_j \epsilon_j} \\
&\ge (a/2)^{\frac{\lambda_j \epsilon_j}{8} \cdot k}
  \ \ \ \text{by the definition of $m_k$, $\overline{m}_k$} \\
&\ge \exp \Big( \frac{\ln(a/2)}{(C')^2 a} \cdot \frac{k}{k^2} \Big)
  \ \ \ \text{by the definition of $\lambda_j$ and $\epsilon_j$} \\
&= e^{c(a)/k}.
\end{align*}
Therefore Equation~\eqref{eqn:beta needed} holds for $\gamma =
c(a)/k$, and this choice of $\gamma$ satisfies the required condition
in Equation~\eqref{eqn:gamma}, since $T = C(a) (k+1) \log(d)$.
\end{proof}
This completes the proof of Lemma~\ref{lem:internal-noise}.
\end{proof}

\subsection{Pruning}\label{sec:pruning}

To this point, we have shown that the Chaining Pursuit Proper
algorithm produces an approximation $\widehat{f}$ of at most $O(m)$
terms with
\[
\smnorm{1}{ f - \widehat{f} }
	\leq (1 + C \log m) \smnorm{1}{ f - f_m }.
\]
We now show that pruning produces $\widehat{f}_m$ with
\[
\smnorm{1}{ f - \widehat{f}_m }
	\leq 3(1 + C \log m) \smnorm{1}{ f - f_m }.
\]
That is, we reduce the number of terms to exactly $m$ while increasing
the error by a small constant factor.  This result applies to any
approximation, not just an approximation produced by Chaining Pursuit
Proper.  This is our top-level result.
\begin{thm}
Let $\widehat{f}$ be an approximation to $f$ with
$\smnorm{1}{ f - \widehat{f} } \leq B\smnorm{1}{ f - f_m }$.  Then
\[
\smnorm{1}{ f - \widehat{f}_m } \leq (2B+1)\smnorm{1}{ f - f_m }.
\]
\end{thm}
\begin{proof}
We have, using the triangle inequality and optimality of
$\widehat{f}_m$ for $\widehat{f}$,
\begin{eqnarray*}
\smnorm{1}{ f - \widehat{f}_m }
& \le & \smnorm{1}{ f - \widehat{f} } 
            + \smnorm{1}{ \widehat {f} - \widehat{f}_m }\\
& \le & \smnorm{1}{ f - \widehat{f} }
            + \smnorm{1}{ \widehat {f} - f_m }\\
& \le & \smnorm{1}{ f - \widehat{f} }
       + \smnorm{1}{ \widehat {f} - f } + \smnorm{1}{ f - f_m }\\
& \le & (2B+1)\smnorm{1}{ f - f_m }.
\end{eqnarray*}
\end{proof}

\subsection{Robustness}

In this subsection, we prove Corollary~\ref{rem:stable}.  As
advertised in the introduction, the Chaining Pursuit algorithm is not
only stable with respect to noise in the signal but also robust to
inaccuracy or errors in the measurements.  Suppose that instead of
using the sketch $\mtx{\Phi}f$ of the signal $f$, e receive $V =
\mtx{\Phi} f + y$ and we reconstruct $\widehat f$ from $V$.  We
assume that once we carry out the Chaining Pursuit algorithm, there
are no perturbations to the intermediate measurements, only to the
original sketch $\mtx{\Phi}f$.

\begin{cor}
With probability at least $(1 - O(d^{-3}))$, the random measurement
operater $\mtx{\Phi}$ has the following property.  Suppose that $f$ is
a $d$-dimensional signal whose best $m$-term approximation with
respect to the $\ell_1$ norm is $f_m$.  Given the measurement operator
$\mtx{\Phi}$, for every $V$ (not necessarily the sketch
$\mtx{\Phi}f$ of $f$), if $\widehat f$ is the reconstruction from
$V$, then
\[   \|f - f_m\|_1 \leq C (1 + \log(m))
           \Big(\|f - f_m\|_1 + \| \mtx{\Phi}f - V\|_1\Big).
\]
\end{cor}
\begin{proof}
We need only make a few adjustments to the proof of the main theorem
to obtain this result.  For brevity, we note these changes.  Let $V =
\mtx{\Phi}f + y$ and let us refer to $y$ as the measurement error.

First, we normalize the signal so that the measurement error has
$\ell_1$ norm $\|y\|_1 = \frac{1}{(800,000 a)}$ and the noise $w = f -
f_m$ has $\ell_1$ norm $\|w\|_1 = \frac{1}{(800,000a)}$.  Next, we
modify the Chaining Recovery conditions for Isolation Matrices in
Condition~\ref{cond:isolation-matrix} to include a third property.
\begin{cond}[Chaining Recovery Conditions for Robust Isolation Matrices]
  \label{cond:isolation-matrix}
  A 0-1 matrix with pass/trial hierarchical structure described in
  Section~\ref{sec:isolation} ({\em i.e.}, any matrix from the sample
  space described in Section~\ref{sec:isolation}) is said to satisfy
  the \term{Chaining Recovery Conditions} if for any signal of the
  form in Invariant~\ref{inv:inductive-hypoth} and for any pass $k$,
  then at least 99/100 of the trial submatrices have these two
  properties:
\begin{enumerate}
\item All but $\tfrac{1}{100} m_{k+1}$ spikes appear alone in a
  measurement, isolated from the other spikes.
\item Except for at most $\tfrac{1}{100} m_{k+1}$ of the measurements,
the internal and external noise assigned to each measurement has
$\ell_1$ norm at most $\tfrac{1}{1000} m_k^{-1}$. 
\item Except for at most $\tfrac{1}{100} m_{k+1}$ of the measurements,
  the measurement error assigned to each measurement has $\ell_1$ norm
  at most $\tfrac{1}{1000} m_k^{-1}$. 
\end{enumerate}
\end{cond}
To prove that a random isolation matrix satisfies this additional
property with high probability, we use Markov's inequality to bound
the number of measurements that are large.  This is the same argument
as in the second half of the proof of Lemma~\ref{lem:external-noise}.

Next, we adjust Lemma~\ref{lemma:simple-bit-test} to include in the
bound $\epsilon$ not just the $\ell_1$ norm of the other positions but
also the measurement error.  We also modify the definition of a good
measurement in Definition~\ref{defn:goodmeas} to include the
measurement error.  
\begin{defn}\label{defn:goodmeas}
A
\term{good measurement} satisfies one of the following two criteria:
\begin{enumerate}
\item The measurement is empty; that is, it contains positions with
  values $\abs{f^{(k)}(i)} \leq \epsilon$ and the total $\ell_1$ norm
  of the positions in the measurement plus the measurement error is
  less than $1.5 \epsilon$.
\item The measurement contains one spike at position $i$ with
  $\abs{f^{(k)}(i)} > \epsilon$ and the $\ell_1$ norm of all other
  positions in this measurement plus the measurement error is less
  than $0.5\epsilon$.
\end{enumerate}
\end{defn}

We conclude by noting that with the above changes, we change
Lemma~\ref{lem:rec_error} to include the $\ell_1$ norm of the
measurement error $\|y\|_1$, as well as the noise $\|w\|_1$.  That is,
after the last pass, the error $\|f^{(K+1)}\|_1$ with
\[  \|f^{(K+1)}\|_1 \leq \|w\|_1 + \Big\| \sum_{j=0}^K \nu_j \Big\|_1 
    \leq \|w\|_1 + \sum_{j=1}^K 6 m_j m_J^{-1} = \|w\|_1 + 6 \log_a m.
\]
Since $a$ is a constant and $\|w\|_1$ and $\|y\|_1$ were normalized to
be constant, we have that the overall error is at most 
\[   (1 + C \log(m)) \Big( \|f - f_m\|_1 + \|\mtx{\Phi} - V\|_1 \Big).
\]
\end{proof}

\section{Algorithmic Dimension Reduction} \label{sec:embed-analysis}

The following dimension reduction
theorem holds for sparse vectors.
\begin{thm}
Let $X$ be the union of all $m$-sparse signals in $\mathbb{R}^d$ and
endow $\mathbb{R}^d$ with the $\ell_1$ norm.  The linear map
$\bm{\Phi}: \mathbb{R}^d \to \mathbb{R}^n$ in
Theorem~\ref{thm:main-result} satisfies
\[  A \|f - g\|_1 \leq \| \bm{\Phi}(f) - \bm{\Phi}(g)\|_1 
    \leq B \|f -g\|_1
\]
for all $f$ and $g$ in $X$, where $1/A = C \log(m)$ and $B
= C\log^2(m) \log^2(d)$ and $n = O(m \log^2 d)$.
\end{thm}
\begin{proof}
The upper bound is equivalent to saying that the $\ell_1 \to \ell_1$
operator norm satisfies $\|\Fee\|_{1 \to 1} \le B$.  This norm is
attained at an extreme point of the unit ball of $\ell_1^d$, which is
thus at a point with support $1$. Then the upper bound follows at once from the
definition of $\Fee$.  That is, any 0-1 vector of support $1$
gets mapped by $\Fee$ to a 0-1 vector of support bounded by the
total number of bit-tests in all trials and passes, which is
$\sum_{k=0}^{\log_a m} O(k \log d) \log_2 d \le B$.

The lower bound follows from Theorem~\ref{thm:main-result}.  Let $f$
and $g$ be any $d$-dimensional signals with support $m$, so that
$f=f_m$ and $g=g_m$.
Let $V = \Fee g$. Then the reconstruction
$\widehat{f}$ from $V$ will be exact: $\widehat{f} = g$.
As proven in Corollary~\ref{rem:stable}, 
\begin{eqnarray*}
\|f - g\|_1
&  =  & \|f - \widehat{f}\|_1\\
& \le &  C \log(m) \left( \smnorm{1}{f-f_m} + \smnorm{1}{\Fee f - V}\right)\\
&  =  &  C \log(m) \; \smnorm{1}{\Fee f - \Fee g},
\end{eqnarray*}
which completes the proof.
\end{proof}

We are interested not only in the distortion and dimension reduction
properties of our embedding but also in the stability and robustness
properties of the embedding.  Our previous analysis guarantees that
$\bm{\Phi}^{-1} \bm{\Phi}$ is the identity on $X$ and that the inverse
can be computed in sublinear time since Chaining Pursuit Proper perfectly
recovers $m$-sparse signals.  Our previous analysis also shows that
our dimension reduction is stable and robust.  In other words, our
embedding and the reconstruction algorithm can tolerate errors $\eta$
in the data $x \in X$, as well as errors $\nu$ in the measurements:

\begin{thm}
  The linear map $\Fee : \mathbb{R}^d \to \mathbb{R}^n$ in Theorem 2
  and the reconstruction map $\mathbf{\Psi} : \mathbb{R}^n \to
  \mathbb{R}^d$ given by the Chaining Pursuit Proper algorithm satisfy the
  following for every $\eta \in \mathbb{R}^d$ and every $\nu \in
  \mathbb{R}^n$ and for all $m$-sparse signals $x$ in $\mathbb{R}^d$:
 $$
 \|x - \mathbf{\Psi}(\Fee(x+\eta)+\nu) \|_1
 \le (1 + \mathrm{C} \log m) (\|\eta\|_1 + \|\nu\|_1).
 $$
\end{thm}
\begin{proof}
  This is just a reformulation of our observations in
  Corollary~\ref{rem:stable} with
  $x = f_m$, $\eta = f-f_m$, $\nu = \Fee f - V$.
\end{proof}

\section{Conclusions}

We have presented the first algorithm for recovery of a noisy sparse
vector from a nearly optimal number of non-adaptive linear
measurements that satisfies the following two desired properties:
\begin{itemize}
\item A single uniform measurement matrix works simultaneously for all
  signals.
\item The recovery time is, up to log factors, proportional to the
  size of the {\em output}, not the length of the vector.
\end{itemize}

The output of our algorithm has error with $\ell_1$-norm bounded in
terms of the $\ell_1$-norm of the optimal output.  Elsewhere in the
literature, {\em e.g.},
in~\cite{CT04:Near-Optimal,RV06:Sparse-Reconstruction,CDD06:Remarks-Compressed},
the $\ell_2$-norm of the output error is bounded in terms of the
$\ell_1$-norm of the optimal error, a mixed-norm guarantee that is
somewhat stronger than the result we give here.  A companion paper, in
progress, addresses this as well as the logarithmic factor in the
approximation error that we give here.

If the measurement matrix is a random Gaussian matrix, as
in~\cite{CT04:Near-Optimal,RV06:Sparse-Reconstruction,CDD06:Remarks-Compressed},
the measurement matrix distribution is invariant under unitary
transformations.  It follows that such algorithms support recovery of
signals that are sparse in a basis {\em unknown at measurement time}.
That is, one can measure a signal $f^*$ as $V=\Fee f^*$.  Later, one
can decide that $f^*$ can be written as $f^*=Sf$, where $S$ is an
arbitrary unitary matrix independent of $\Fee$ and $f$ is a noisy
sparse vector of the form discussed above.  Thus $V=(\Fee S)f$, where
$\Fee S$ is Gaussian, of the type required by the recovery algorithm.
Thus, given $V,\Fee$, and $S$, the algorithms
of~\cite{CT04:Near-Optimal,RV06:Sparse-Reconstruction,CDD06:Remarks-Compressed}
can recover $f$.

If the matrix $S$ is known at measurement time, our algorithm can
substitute $\Fee S$ for $\Fee$ {\em at measurement time} and proceed
without further changes.  If $S$ is unknown at measurement time,
however, our algorithm breaks down.  But note that an important
point of our algorithm is to provide decoding in time $m\polylog(d)$,
which is clearly not possible if the decoding process must first read
an arbitrary unitary $d$-by-$d$ matrix $S$.  Once a proper problem has
been formulated, it remains interesting and open whether
sublinear-time decoding is compatible with basis of sparsity unknown
at measurement time.

\bibliographystyle{alpha}
\bibliography{stoc2006}

\end{document}